\documentclass[twocolumn,apj,iop,tighten]{emulateapj2}

\usepackage{float}
\usepackage{graphicx}
\usepackage{xfrac}
\usepackage{adjustbox}

\usepackage{natbib}
\usepackage[usenames]{color}

\definecolor{black}{rgb}{0.0, 0.0, 0.0}
\definecolor{darkblue}{rgb}{0.0, 0.0, 0.55}
\definecolor{blue}{rgb}{0.0, 0.0, 1.0}
\definecolor{dodgerblue}{rgb}{0.12, 0.56, 1.0}
\definecolor{cyan}{rgb}{0.0, 1.0, 1.0}
\definecolor{lime}{rgb}{0.0, 1.0, 0.0}
\definecolor{green}{rgb}{0.0, 0.5, 0.0}
\definecolor{red}{rgb}{1.0, 0.0, 0.0}
\definecolor{salmon}{rgb}{1.0, 0.6, 0.6}

\definecolor{purple}{rgb}{0.5411764705882353, 0.16862745098039217, 0.8862745098039215}
\definecolor{magenta}{rgb}{1.0, 0.0, 1.0}

\usepackage[ colorlinks=true, 
             urlcolor=blue,
             filecolor=blue,
             linkcolor=red,
             citecolor=blue,
             pdftitle={pdf title},
             pdfauthor={pdf author},
             pagebackref, 
             bookmarks,
             bookmarksopen=true]{hyperref}

\usepackage{afterpage}
\usepackage[caption=false]{subfig}
\usepackage{longtable}

\usepackage{xspace}    
\usepackage{microtype} 
\usepackage{amsmath}   
\usepackage{amssymb}

\usepackage[normalem]{ulem}
\newcommand\redline{\bgroup\markoverwith{\textcolor{red}{\rule[0.5ex]{2pt}{0.4pt}}}\ULon}

\usepackage{multirow}
\usepackage{makecell} 
\usepackage{rotating} 

\newcommand{\degsq}{\ensuremath{\mathrm{deg}^2}\xspace}
\newcommand{\dvmax}{\ensuremath{\Delta\vmax}\xspace}

\newcommand{\halotools}{\texttt{Halotools}\xspace}

\newcommand{\iSEDfit}{\texttt{iSEDfit}\xspace}

\newcommand{\logm}{\ensuremath{\log(M_*/\msun)}\xspace}

\newcommand{\logssfr}{\ensuremath{\log({\rm sSFR}/{\rm yr}^{-1})}\xspace}

\newcommand{\mstar}{\ensuremath{M_*}\xspace}
\newcommand{\msun}{\ensuremath{M_\odot}\xspace}
\newcommand{\Mpch}{\ensuremath{h^{-1}\,{\rm Mpc}}\xspace}

\newcommand{\onehalo}{\ensuremath{0.1~\Mpch<\rp<1~\Mpch}\xspace}

\newcommand{\pimax}{\ensuremath{\pi_{\rm max}}\xspace}

\newcommand{\rockstar}{\textsc{rockstar}\xspace}
\newcommand{\rp}{\ensuremath{r_{\rm p}}\xspace}

\newcommand{\sigmaz}{\ensuremath{\sigma_z}\xspace}

\newcommand{\twohalo}{\ensuremath{1~\Mpch<\rp<10~\Mpch}\xspace}

\newcommand{\UM}{\textsc{UniverseMachine}\xspace}

\newcommand{\vmax}{\ensuremath{V_{\rm max}}\xspace}

\renewcommand{\wp}{\ensuremath{\omega_{\rm p}}\xspace}
\newcommand{\wprp}{\ensuremath{\omega_{\rm p}(r_{\rm p})}\xspace}

\newcommand{\xir}{\ensuremath{\xi(r)}\xspace}
\newcommand{\xirppi}{\ensuremath{\xi(r_{\rm p},\,\pi)}\xspace}

\newcommand{\zhigh}{\ensuremath{0.7<z<1.2}\xspace}
\newcommand{\zlow}{\ensuremath{0.2<z<0.7}\xspace}

\newcommand{\zquality}{\ensuremath{z_{\rm quality}}\xspace}

\begin{document}
\title{Main Sequence Scatter is Real:\ The Joint Dependence of Galaxy Clustering on Star Formation and Stellar Mass}
\shortauthors{Berti et al.}

\author{Angela M.\ Berti,\altaffilmark{1}
	Alison L.\ Coil,\altaffilmark{1}
	Andrew P.\ Hearin,\altaffilmark{2}
	Peter S.\ Behroozi\altaffilmark{3}
	}
	
\altaffiltext{1}{Center for Astrophysics and Space Sciences, Department of Physics, University of California, 9500 Gilman Dr., La Jolla, San Diego, CA 92093, USA}
\altaffiltext{2}{High-Energy Physics Division, Argonne National Laboratory, Argonne, IL 60439, USA}
\altaffiltext{3}{Department of Astronomy and Steward Observatory, University of Arizona, 933 N Cherry Ave., Tucson, AZ 85719, USA}

\begin{abstract}
We present new measurements of the clustering of stellar mass-complete samples of $\sim40,000$ SDSS galaxies at $z\sim0.03$ as a joint function of stellar mass and specific star formation rate (sSFR). Our results confirm what \citet{coil_etal17} find at $z\sim0.7$:\ galaxy clustering is a stronger function of sSFR at fixed stellar mass than of stellar mass at fixed sSFR. We also find that galaxies above the star-forming main sequence (SFMS) with higher sSFR are less clustered than galaxies below the SFMS with lower sSFR, at a given stellar mass. A similar trend is present for quiescent galaxies. This confirms that main sequence scatter, and scatter within the quiescent sequence, is physically connected to the large-scale cosmic density field.
We compare the resulting galaxy bias versus sSFR, and \emph{relative} bias versus sSFR \emph{ratio}, for different galaxy samples across ${0<z<1.2}$ to mock galaxy catalogs based on the empirical galaxy evolution model of \citet{behroozi_etal19}. This model fits PRIMUS and DEEP2 clustering data well at intermediate redshift, but agreement with SDSS is not as strong. We show that increasing the correlation between galaxy SFR and halo accretion rate at $z\sim0$ in the model substantially improves agreement with SDSS data.
Mock catalogs suggest that central galaxies contribute substantially to the dependence of clustering on sSFR at a given stellar mass and that the signal is not simply an effect of satellite galaxy fraction differences with sSFR.
Our results are highly constraining for galaxy evolution models and show that the stellar-to-halo mass relation (SHMR) depends on sSFR.
\end{abstract}

\section{Introduction}\label{sec04:intro}

In the $\Lambda$CDM paradigm, galaxies form at the centers of collapsing overdensities in a cosmic web of dark matter that underlies the large-scale structure of the universe.
Large $N$-body cosmological simulations model the predicted evolution of the structure of dark matter, while galaxy redshift surveys reveal the spatial distribution of observed galaxies and provide observational constraints for models of the galaxy--halo connection.

Theoretical models for linking galaxies to halos typically use either stellar mass or luminosity as the primary determining characteristic of galaxy clustering \citep[e.g.][]{kravtsov_etal04, vale_ostriker04, vale_ostriker06}.
These models also usually differentiate between central galaxies (primary galaxies at the centers of their halos) and satellites galaxies (residing in subhalos within a larger, more massive parent halo). Although it is often not clear within observed galaxies populations if an individual galaxy is a central or satellite, this distinction has proven useful for galaxy--halo connection models because, while centrals and satellites can occupy the same parent halo, they have different formation and evolutionary histories.

The best-fit parameters in galaxy--halo models are different for galaxy populations split by properties such as luminosity and optical color.
This offers insight into the physical processes responsible for the observed clustering properties of galaxies, which are the main constraints on models of the galaxy--halo connection.
In particular, the observed dependence of galaxy clustering on properties such as luminosity, stellar mass, and color have been thoroughly studied within the limits of existing survey data to $z\sim1$.
Clustering amplitude positively correlates with luminosity, particularly for $L>L^*$, while the correlation is shallower for fainter galaxies \citep{benoist_etal96, norberg_etal01, coil_etal06, pollo_etal06, meneux_etal09}.
Similar, although weaker, trends are observed with stellar mass, particularly for masses greater than $\sim M^*$ \citep{li_etal06, meneux_etal08, wake_etal11, leauthaud_etal12, marulli_etal13}.
Coupled with the known dependence of halo clustering amplitude on halo mass, this result implies a stellar-to-halo mass relation (SHMR), estimates of which at different redshifts provide insight into the evolution of star formation efficiency and galaxy evolution \citep{behroozi_etal10, moster_etal10, leauthaud_etal11, durkalec_etal15, skibba_etal15}.

Galaxy clustering studies have more recently begun to explore the dependence on star formation rate (SFR) and specific SFR (sSFR, or SFR per unit stellar mass), which is more closely linked to the physical processes relevant for star formation than optical color.
\citet{heinis_etal09} divide star-forming SDSS galaxies into two bins in sSFR and find that clustering amplitude increases with decreasing sSFR.
At higher redshift \citet{mostek_etal13} find with data from the DEEP2 survey that clustering amplitude is stronger with increasing SFR and decreasing sSFR, although they acknowledge these trends are largely but not entirely attributable to the correlation between SFR and stellar mass, i.e.\ the star-forming main sequence (SFMS).
\citet{mostek_etal13} also find that within the star-forming population galaxies above the main sequence are less clustered than those below for a given stellar mass range, implying that galaxies evolve not only \emph{along} the SFMS as they build up stellar mass, but also \emph{across} it, from above to below.

\citet[][hereafter C17]{coil_etal17} use the PRIMUS and DEEP2 datasets to further subdivide galaxies in the stellar mass--sSFR plane at $z\sim0.5$ and $z\sim0.9$, and find that galaxy clustering depends as strongly on sSFR as on stellar mass.
Specifically, C17 find a strong correlation between the relative clustering amplitudes of different galaxy samples and the ratio of their sSFRs at a given stellar mass ratio, but not vice versa:\ relative clustering strength shows little dependence on stellar mass ratio at fixed sSFR ratio.
This indicates that stellar mass may not be the primary galaxy property relevant for clustering and the galaxy--halo connection.

Complimenting observations of galaxy clustering dependencies across cosmic time are studies of halo clustering properties performed with $\Lambda$CDM cosmological simulations.
\emph{Halo assembly bias} refers to the finding that halo clustering depends on factors beyond halo mass, including halo age \citep[e.g.][]{gao_etal05} and concentration \citep{wechsler_etal06, villarreal_etal17}, among others \citep{dalal_etal08, mao_etal18, salcedo_etal18, johnson_etal19, mansfield_kravtsov20}.
Combined with the observed clustering dependencies of galaxy properties, this has led to hypotheses of \emph{galaxy assembly bias}, the correlation of a secondary galaxy property (other than stellar mass), such as luminosity or star formation rate, with an additional tracer of halo assembly history, such as dark matter accretion rate or maximum circular velocity \vmax \citep[e.g.][]{wechsler_tinker18, zentner_etal19}.

While halo assembly bias is a well-established prediction of $\Lambda$CDM simulations, the role of galaxy assembly bias in the relationship between galaxies and halos is an open question. The direct dependence of galaxy properties on halo properties beyond mass is inconsistent with the assumption that a halo's mass is sufficient to statistically predict its galaxy content, but definitive observational evidence is difficult to come by, especially at higher redshift.
For example, C17's result that clustering does \emph{not} depend on stellar mass \emph{at a given sSFR} is consistent with the SHMR depending on galaxy sSFR, which if true would be a manifestation of galaxy assembly bias. However, C17 note they cannot eliminate the possibility their results are due to satellite galaxies, which are known to reside in more massive halos than central galaxies of the same stellar mass \citep{watson_conroy13} and also have a larger quiescent fraction than central galaxies in less massive halos \citep{wetzel_etal12}.

Using ``isolated primary" galaxies in PRIMUS data as a proxy for centrals, \citet{berti_etal19} investigate the joint clustering dependence of central galaxies on stellar mass and sSFR at $z\sim0.35$ and $z\sim0.7$. They compare their results to mock galaxy catalogs based on the empirical \UM model of \citet{behroozi_etal19} and find that C17's results for all galaxies also hold for centrals:\ quiescent central galaxies are significantly more clustered than star-forming centrals at fixed stellar mass. This is consistent with some combination of central galaxy assembly bias and distinct SHMRs for quiescent and star-forming central galaxies.

Observationally, the distinction between centrals and satellites cannot be drawn as cleanly as is possible with dark matter $N$-body simulations. Existing methods for distinguishing central galaxies from satellites in the data---such as group-finding algorithms, isolation criteria, and using proxies like brightest cluster galaxies---are each subject to systematic uncertainties that manifest as varying levels of sample incompleteness (missing true centrals) and contamination (misclassifying a satellite galaxy as a central).
This makes it difficult to directly compare theoretical models that divide galaxies into centrals and satellites with observational data.
What is possible is to determine existing observational constraints for the entire galaxy population via empirical forward modeling, and see what novel implications about galaxy evolution---including correlations between properties of galaxies and halos---emerge from a particular model \citep[e.g.][]{behroozi_etal13, behroozi_etal15}. Properties of central and satellites galaxy populations can be separately constrained in this way as well.

In this paper we measure the joint dependence of galaxy clustering on stellar mass and sSFR at $z\sim0$ using data from the tenth SDSS data release, extending C17's study of this joint dependence at $0.2 < z < 1.2$ to the local universe for the first time. 
We compare our and C17's measurements to mock galaxy catalogs created using the \UM model of \citet{behroozi_etal19} to test its agreement with observations---specifically clustering amplitude differences with sSFR within both the star-forming and quiescent populations, or \emph{intra-sequence relative bias} (ISRB)---from $z\sim0$ to $z\sim1$.
While we find strong agreement between the data and model at higher redshift, we demonstrate how the model can be updated to better fit SDSS data.
We then use the simulations to assess the relative contributions of central and satellite galaxies to the ISRB observed at both $z=0$ and to $z=1$.

The structure of this paper is as follows.
In \S\ref{sec04:data_sims} we provide an overview of the data and mock galaxy catalogs we use, as well as the galaxy samples used for our clustering measurements.
\S\ref{sec04:methods} describes our methods for measuring clustering amplitudes, absolute and relative biases, and estimating errors.
In \S\ref{sec04:wp_bias} we present our measurements of galaxy clustering as a joint function of sSFR and stellar mass at $z\sim0$, and compare these results and C17's analogous measurements at higher redshift to mock catalogs.
\S\ref{sec04:z0_mod} describes how we modify \UM model at $z=0$ to improve agreement with observations, and in \S\ref{sec04:cen_sat} we use simulations to investigate the sources of intra-sequence clustering amplitude differences.
We summarize our results in \S\ref{sec04:summary}.
Throughout this paper we assume a standard $\Lambda{\rm CDM}$ cosmology with $H_0=70$~km~s$^{-1}$~Mpc$^{-1}$. The cosmological parameters of the simulations used are given in \S\ref{subsec04:sims_mocks}.

\section{Data and Simulations}\label{sec04:data_sims}

In this section we describe the observational datasets, $N$-body dark matter simulations, and mock galaxy catalogs used in this study. We use data from the Sloan Digital Sky Survey to report new galaxy clustering measurements at $z\sim0$. We further use simulations to create mock catalogs to compare with these new $z\sim0$ clustering measurements, as well as previously-published clustering measurements from the PRIMUS and DEEP2 galaxy redshift surveys at $z\sim0.45$ and $z\sim0.9$.

\subsection{SDSS}\label{subsec04:sdss}

We use galaxy redshifts from Data Release 10 of the Sloan Digital Sky Survey \citep[SDSS;][]{ahn_etal14}. Stellar mass and SFR measurements are taken from the MPA-JHU catalog \citep{kauffmann_etal03, brinchmann_etal04}. In this catalog, fiber SFRs are measured from H$\alpha$ (for star-forming galaxies) and estimated from the D4000 break (for quiescent galaxies) for the light within the SDSS fiber.
Light outside each galaxy's fiber is converted to an SFR assuming the same average SFR/luminosity ratio as other fibers with similar $g-r$ and $r-i$ colors.
The total SFRs used here are the sum of the fiber and non-fiber SFRs.

To create stellar mass complete samples we use a modified version of the redshift-dependent $r$-band apparent magnitude completeness cut of \citet{behroozi_etal15}:
\begin{equation}\label{eq04:sdss_mass_comp}
    r<-0.25-1.9\,\log\left[\frac{\mstar}{\msun}\right]+5\,\log\left[\frac{D_{\rm L}(z)}{10\,{\rm pc}}\right],
\end{equation}
\noindent where $D_{\rm L}(z)$ is the luminosity distance.
\citet{behroozi_etal15} find that at least 96\% of SDSS galaxies within {$9.5<\logm<10.0$} satisfy {$M_r<-0.25-1.9\logm$}, where $M_r$ is the galaxy's Petrosian $r$-band absolute magnitude. This completeness limit becomes Eq.\ \ref{eq04:sdss_mass_comp} when expressed in terms of redshift and apparent $r$-band magnitude. Inverting Equation~\ref{eq04:sdss_mass_comp} with $r=17.77$ gives the minimum stellar mass to which SDSS is $\gtrsim96\%$ complete as a function of redshift.
Our SDSS samples are complete to a minimum stellar mass of $\mstar>10^{9.75}\ \msun$ at $z<0.0435$, which we use as the maximum redshift of our SDSS galaxy samples.

We note, however, that bluer, star-forming galaxies have greater mass-to-light ratios than redder, quiescent galaxies of the same stellar mass. Thus while $10^{9.75}\ \msun$ is an appropriate mass completeness limit for quiescent SDSS galaxies at $z<0.0435$, this cut is unnecessarily conservative for star-forming galaxies.
Our goal in this study is to probe as wide of a range of galaxy stellar mass as possible, while maintaining an adequately large sample size for robust statistics.
Therefore we determine a lower stellar mass above which star-forming SDSS galaxy samples will be complete.

First we classify galaxies as star-forming or quiescent based on whether they fall above or below the following redshift-dependent cut in the stellar mass--sSFR plane:
\begin{equation}\label{eq04:sdss_ssfr_cut}
    \log\left[\frac{{\rm sSFR}}{{\rm yr}^{-1}}\right] = (0.37\,z - 0.48)\log \left[\frac{\mstar}{\msun}\right] + 3.45\,z - 6.10.
\end{equation}
We obtain this cut by dividing the full SDSS sample into narrow redshift bins containing roughly equal numbers of galaxies, and plotting the stellar mass--sSFR distribution in each redshift bin.
Next we fit a line of the form ${\rm sSFR}=\alpha\,\log(\mstar/\msun) + \beta$ to the star-forming main sequence (SFMS) in each redshift bin, and then shift this line downward in sSFR to intersect the minimum of the bimodal galaxy stellar mass--sSFR distribution in that bin.
We then obtain linear fits to the slope $\alpha$ and intercept $\beta$ versus the median redshift in each bin to estimate the redshift dependence of each, and substitute in these linear expressions for $\alpha(z)$ and $\beta(z)$ to obtain Equation~\ref{eq04:sdss_ssfr_cut}.

To determine an appropriate stellar mass completeness cut for star-forming galaxies we measure in narrow redshift bins the fraction of quiescent galaxies above the \citet{behroozi_etal15} stellar mass limit at the median redshift of each bin.
We then identify the stellar mass $M_{\rm min}^{\rm SF}(z)$ at which the same fraction of star-forming galaxies in each bin satisfies $\mstar \geq M_{\rm min}^{\rm SF}(z)$.
We find that star-forming galaxies have the same completeness fraction as their quiescent counterparts at stellar masses 0.5--0.6 dex less than the quiescent galaxy limit of $10^{9.75}\ \msun$ over the redshift range of our sample, with a mean value of 0.55 dex. We therefore adopt a stellar mass completeness limit for star-forming galaxies of $10^{9.25}\ \msun$ at $z=0.0435$.

Our resulting stellar mass complete SDSS sample contains 41,486 star-forming galaxies with $\mstar\geq10^{9.25}\ \msun$ and 17,960 quiescent galaxies with $\mstar\geq10^{9.75}\ \msun$ in the redshift range $0.02<z<0.0435$.

\subsection{PRIMUS and DEEP2}\label{subsec04:primus}

A primary goal of this paper is to compare the sSFR dependence of galaxy clustering in data and mock catalogs at $0<z<1.2$. 
C17 has measured this dependence at $0.2<z<1.2$ using data from both the PRIsm- MUlti-object Survey \citep[PRIMUS][]{coil_etal11, cool_etal13} and DEEP2 \citep{newman_etal13} spectroscopic galaxy redshift surveys.

We refer the reader to C17 for full details of the PRIMUS and DEEP2 galaxy samples used in that work---which are the basis for the SDSS and mock galaxy samples used here (see \S\ref{subsec04:samples} below)---as well as for additional details about each survey.
Briefly, PRIMUS is a low-resolution ($R\sim40$) spectroscopic redshift survey covering $\sim9$~\degsq in seven fields.
Conducted with the IMACS instrument \citep{bigelow_dressler03} on the Magellan I Baade 6.5~m telescope, PRIMUS is the largest spectroscopic faint galaxy redshift survey completed to date.
The survey utilized targeting weights to achieve a statistically complete sample of $\sim120,000$ robust spectroscopic redshifts.

The DEEP2 survey was conducted with the DEIMOS spectrograph \citep{faber_etal03} on the 10\ m Keck II telescope, and contains $\sim17,000$ high-confidence redshifts \citep[$<95\%$ or $Q \geq 3$; see][]{newman_etal13}.

The galaxy samples used in Coil+17 consist of robust \citep[$\zquality \geq 3$; see][]{coil_etal11} PRIMUS redshifts from the CDFS-SWIRE, COSMOS, ES1, and XMM-LSS fields, augmented with $Q\geq3$ DEEP2 redshifts at $0.2<z<1.2$ from the EGS field. For the remainder of this paper ``PRIMUS data" refers to this combined dataset using the PRIMUS and DEEP2 surveys.

C17 estimate stellar masses and SFRs for PRIMUS galaxies by using \iSEDfit \citep{moustakas_etal13} to fit spectral energy distributions (SEDs) with the population synthesis models of \citet{bruzual_charlot03}, assuming a \citet{chabrier03} initial mass function from 0.1 to 100 \msun. We refer the reader to \S2.3 of C17 and to \citet{moustakas_etal13} for additional details about PRIMUS stellar mass and SFR estimates.

\subsection{Simulations and Mocks}\label{subsec04:sims_mocks}

As the basis for mock galaxy catalogs to compare to observational data we use snapshots of dark matter $N$-body simulations at the median redshift of the corresponding data sample.

For $z=0$ we use the Bolshoi simulation\footnote{https://www.cosmosim.org/cms/simulations/bolshoi/} \citep{klypin_etal11}, which contains $2048^3$ particles in a $(250 \Mpch)^3$ cubic volume, has a dark matter particle mass resolution of $1.35\times10^8\ \msun\ h^{-1}$, and uses $\Omega_{\rm m}=0.27$, $\Omega_{\Lambda}=0.73$, and $\sigma_8=0.82$.
For $z=0.45$ and $z=0.9$ we use the Bolshoi-Planck simulation \citep{klypin_etal16}, which is similar to the Boishoi simulation but uses Planck cosmological parameters of $\Omega_{\rm m}=0.31$ and $\Omega_{\Lambda}=0.69$.

We use the publicly-available \UM \citep{behroozi_etal19} code to populate simulation snapshots at $z=0$, $0.45$, and $0.9$ with synthetic galaxies to create mock galaxy catalogs to which we compare our observational results.
In each snapshot (sub)halos are identified with the publicly available \rockstar halo finder code \citep{behroozi_etal13}.
\UM empirically models the dependence of galaxy SFR as a function of halo mass, halo accretion rate, and redshift to predict the star formation histories (SFHs) of individual galaxies over cosmic time and connect those SFHs to the assembly histories of dark matter halos.
Halos are populated with synthetic galaxies based on the distributions of a variety of observed galaxy properties across redshifts from $z\sim0$ to $z\sim10$ \citep[summarized in Table~1 of][]{behroozi_etal19}.

Comparing observations with simulations requires mock galaxy catalogs that have the same joint stellar mass and sSFR distributions, galaxy number density, and line-of-sight positional uncertainty (analogous to redshift error in observational data) as the dataset of interest.
To achieve this we use a procedure similar to the one described in \S3.1 of \citet{berti_etal19}.
Briefly, for each dataset we create a 2D kernel density estimate (KDE) of the joint distribution of galaxy stellar mass versus sSFR.
We then select from the relevant synthetic galaxy population a sample with the same 2D distribution as the corresponding observational data.
Next we randomly down-sample the selected synthetic galaxy population such that it has the same number density as that of the relevant dataset at its median redshift, i.e.\ we match the number density of the $z=0$ snapshot to our SDSS sample, and of the $z=0.45$ and $z=0.9$ snapshots to the number density of PRIMUS data at $0.4<z<0.5$ and $0.85<z<0.95$, respectively.

Finally, we add noise to the line-of-sight coordinate $r_z$ of all galaxies in the $z=0.45$ and $z=0.9$ mocks to simulate the redshift errors of the PRIMUS dataset.
Specifically, we define a ``noisy" line-of-sight coordinate $r^{\rm noisy}_z = r_z + \Delta r_z$, where $\Delta r_z$ is drawn from a normal distribution with a dispersion equal to the distance in \Mpch equivalent to the PRIMUS redshift error $(\sigmaz/(1+z)\approx0.0033)$ at the redshift of the mock.
At $z=0.45$ ${\sigma_{r_z}\simeq16.4}$ \Mpch and at $z=0.9$
${\sigma_{r_z}\simeq21.5}$ \Mpch.
Redshift errors in SDSS are sufficiently small \citep{york_etal00, blanton_etal05} that we do not add additional line-of-sight position noise to the $z=0$ mock.

We note that \citet{coil_etal11} published estimated PRIMUS redshift errors of $\sigmaz\simeq0.005\,(1+z)$, based on a comparison with higher resolution spectroscopic redshifts. However, \citet{behroozi_etal19} found that $\sigmaz\simeq0.0033\,(1+z)$ is a more accurate estimate of the true PRIMUS redshift errors (see Figure C6 of \citet{behroozi_etal19} and the associated discussion).
We test using both $\sigmaz/(1+z)=0.0033$ and $0.005$ to add line-of-sight position uncertainties to the $z=0.45$ and $z=0.9$ mocks.
At both redshifts the clustering amplitudes and biases of the mocks are in better agreement with PRIMUS data if $\sigmaz/(1+z)=0.0033$ is used; larger values of $\sigmaz/(1+z)=0.005$ systematically lower the observed clustering as mock galaxies are overly scattered along the line-of-sight dimension.

Mock galaxies are divided into star-forming and quiescent populations with the following cuts in the stellar mass--sSFR plane:
\begin{subequations}\label{eq04:mock_ssfr_cut}
\begin{align}
    z=0\phantom{.00} & : \log({\rm sSFR}) > -0.46\log(\mstar) - 6.24; \\
    z=0.45 & : \log({\rm sSFR}) > -0.25\log(\mstar) - 8.06; \\
    z=0.9\phantom{0} & : \log({\rm sSFR}) > -0.19\log(\mstar) - 8.35,
\end{align}
\end{subequations}
\noindent where sSFR is in units of yr$^{-1}$ and \mstar is in units of \msun. These cuts are determined using an analogous method to that described in \S\ref{subsec04:sdss} for SDSS galaxies: finding a linear fit to the SFMS in the stellar mass--sSFR plane, and shifting this line downward in sSFR so that it intersects the minimum of the bimodal galaxy distribution in this plane.

\subsection{Galaxy Samples}\label{subsec04:samples}

Two main goals of this paper are (i) to measure the joint dependence of clustering on stellar mass and sSFR in the SDSS, and (ii) compare these measurements in data from both SDSS and PRIMUS to simulations by making analogous measurements in mock galaxy catalogs at $0<z<1.2$. C17 has measured the clustering dependence of galaxies on stellar mass and sSFR at $0.2<z<1.2$ with data from the PRIMUS and DEEP2 galaxy redshift surveys (``PRIMUS data" as described above), and their galaxy samples are the basis for both our SDSS samples (see Figure~\ref{fig04:sdss_samples_sm9p75q_9p25sf} and Table~\ref{tab04:samples_sdss}) and the mock galaxy samples we create at each redshift (see Figures~\ref{fig04:samples_z0p0_obs} and \ref{fig04:samples_z0p45_obs} and Table~\ref{tab04:samples_mock}). We therefore describe C17's samples and the rationale behind them in some detail. 

C17 divide PRIMUS data into two redshift bins, \zlow and \zhigh. Within each redshift bin they create subsamples in four different ways to conduct distinct ``runs" of their analysis. In their nomenclature, {\bf Run 1 (``star-forming/quiescent split")} compares the clustering of stellar mass complete star-forming and quiescent galaxies within the same stellar mass range.

{\bf Run 2 (``main sequence split")} divides the star-forming and quiescent populations within a given stellar mass range into two subsamples each:\ those above and below the star-forming or quiescent main sequence in the stellar mass--sSFR plane.
The goal with these samples is to compare the clustering of star-forming (or quiescent) galaxies with above average sSFRs to those with below average sSFRs at fixed stellar mass.

C17's {\bf Run 3 (``sSFR cuts")} again limits the star-forming and quiescent populations in each redshift bin to specified ranges in stellar mass, and divides each population into three bins in sSFR.  These samples are therefore defined by strict limits in sSFR. 

Finally, {\bf Run 4} is designed to measure the dependence of clustering on sSFR at fixed stellar mass, as well as the dependence of clustering on stellar mass at fixed sSFR. C17 divide PRIMUS data into nine samples at \zlow and seven samples at \zhigh with multiple cuts in both stellar mass and sSFR to create a grid in the stellar mass--sSFR plane. We refer readers to \S3 in C17 for complete descriptions of the stellar mass, sSFR, and redshift cuts that define their samples.

\subsubsection{SDSS Data Samples}\label{subsubsec04:samples_sdss}

As described above, C17's PRIMUS galaxy sample divisions are the basis for our comparable analysis of SDSS data. As PRIMUS is at higher redshift, only C17's ``star-forming/quiescent split" samples are stellar mass complete. Here all of our SDSS samples (described below) are stellar mass complete, using the SDSS mass completeness limits described in \S\ref{subsec04:sdss} above.

Following C17, we divide the SDSS parent sample four ways, shown in Figure~\ref{fig04:sdss_samples_sm9p75q_9p25sf} and described below. Table~\ref{tab04:samples_sdss} contains parameters for our SDSS galaxy sample cuts.
\begin{enumerate}
    \item
{\bf Star-forming/quiescent split.} Here we divide the parent sample into star-forming and quiescent subsamples, both limited to $9.75<\log(\mstar/\msun)<10.5$, to compare the star-forming and quiescent populations at fixed stellar mass. Galaxies are classified as star-forming or quiescent based on Equation~\ref{eq04:sdss_ssfr_cut} (see \S\ref{subsec04:sdss}).
    \item    
{\bf Main sequence split.} Here we divide the star-forming and quiescent sequences each into two samples. We find the median sSFR of the star-forming population in narrow bins in stellar mass, and split the star-forming galaxies into those above and those below the median sSFR for each stellar mass bin. This method creates two star-forming samples with identical stellar mass distributions, allowing us to compare the clustering of star-forming galaxies with above average verses below average sSFRs at fixed stellar mass, and likewise for the quiescent population. The samples above and below the star-forming main sequence have mean stellar masses of $\logm=9.87$, and respective mean sSFRs of $\logssfr=-9.87$ and $-10.37$. The samples above and below the quiescent sequence have mean stellar masses of $\logm=10.42$, with mean sSFRs of $\logssfr=-11.60$ and $-12.14$, respectively.
    \item
{\bf sSFR cuts.} Here we use cuts in sSFR to divide the parent sample into six subsamples, three spanning the star-forming population and three  spanning the quiescent population. The star-forming samples are  limited to $\logm>9.25$ with sSFR cut lower bounds at ${\logssfr=-10.0}$, $-10.4$, and $-10.8$. The quiescent samples are limited to $\logm>9.75$ with sSFR cut upper bounds at ${\logssfr=-10.8}$, $-11.5$, and $-12.1$. While all of these samples are stellar mass complete, unlike the previous set of samples the mean stellar mass varies by $\sim0.5$ dex within the three star-forming subsamples, from $\sim9.6$ to $\sim10.1$. Within the quiescent subsamples the mean stellar mass ranges from $\sim10.3$ to $\sim10.7$.
    \item
{\bf Stellar mass/sSFR grid.} The final set of subsamples is designed to allow us to measure both the clustering dependence on stellar mass at fixed sSFR, and on sSFR at fixed stellar mass. For these subsamples we divide the stellar mass--sSFR plane into three stellar mass bins with lower bounds at $\logm=9.25$, $9.75$, and $10.4$, and four bins in sSFR with cuts at $\logssfr=-10.0$, $-10.8$, and $-11.7$. Of the 12 regions in the stellar mass--sSFR plane defined by these cuts, nine contain sufficient numbers of galaxies to be included in our analysis. These nine regions are by design are comparable to C17's divisions of PRIMUS data at \zlow, although the precise values of the bin edges are offset from those used at higher redshift to better align with the distribution of SDSS data in the stellar mass--sSFR plane.
\end{enumerate}

\begin{deluxetable*}{llrrrrrrrrr}[h]
\tablecaption{SDSS galaxy samples. All samples span $0.02 < z < 0.0435$ and have median redshift $z_{\rm med}\simeq0.033$.
\label{tab04:samples_sdss}
}

\tablehead{
\colhead{\multirow{2}{*}{Run}} &
\colhead{\multirow{2}{*}{Sample}} &
\colhead{\multirow{2}{*}{$N_{\rm gal}$}} &
\multicolumn{3}{c}{\logm} &
\multicolumn{3}{c}{$\log($sSFR/yr$^{-1})$} &
\multicolumn{2}{c}{Bias\tablenotemark{a}} \\
\colhead{} & \colhead{} & \colhead{} &
\colhead{min} & \colhead{mean} & \colhead{max} &
\colhead{min} & \colhead{mean} & \colhead{max} &
\colhead{one-halo} & \colhead{two-halo} \\
\cline{1-11}
}

\startdata
${\rm SF/Q\ split}$	& \textcolor{blue}{\textbf{blue}}	& 11009	& 9.75	& 10.08	& 10.50	& -11.14	& -10.23	& -8.41	& $0.78\,(0.03)$	& $1.19\,(0.05)$ \\
	& \textcolor{red}{\textbf{red}}	& 7751	& 9.75	& 10.15	& 10.50	& -13.40	& -11.69	& -10.80	& $2.26\,(0.06)$	& $1.97\,(0.06)$ \\
${\rm Main\ sequence\ split}$	& \textcolor{darkblue}{\textbf{dark blue}}	& 12582	& 9.25	& 9.87	& 11.35	& -11.20	& -9.87	& -8.41	& $0.69\,(0.03)$	& $1.08\,(0.05)$ \\
	& \textcolor{dodgerblue}{\textbf{light blue}}	& 12604	& 9.25	& 9.87	& 11.43	& -11.54	& -10.37	& -9.77	& $0.92\,(0.03)$	& $1.33\,(0.05)$ \\
	& \textcolor{salmon}{\textbf{light red}}	& 6712	& 9.75	& 10.42	& 11.59	& -12.70	& -11.60	& -10.80	& $1.63\,(0.05)$	& $1.72\,(0.06)$ \\
	& \textcolor{red}{\textbf{red}}	& 6732	& 9.75	& 10.42	& 11.75	& -13.40	& -12.14	& -11.40	& $2.37\,(0.06)$	& $1.99\,(0.05)$ \\
${\rm sSFR\ cuts}$	& \textbf{black}	& 10695	& 9.25	& 9.63	& 11.15	& -10.00	& -9.76	& -9.00	& $0.67\,(0.03)$	& $1.05\,(0.05)$ \\
	& \textcolor{blue}{\textbf{blue}}	& 8452	& 9.25	& 9.91	& 11.24	& -10.40	& -10.19	& -10.00	& $0.82\,(0.03)$	& $1.22\,(0.05)$ \\
	& \textcolor{dodgerblue}{\textbf{light blue}}	& 4879	& 9.25	& 10.09	& 11.35	& -10.80	& -10.58	& -10.40	& $1.11\,(0.04)$	& $1.49\,(0.05)$ \\
	& \textcolor{lime}{\textbf{light green}}	& 4342	& 9.75	& 10.27	& 11.57	& -11.50	& -11.16	& -10.80	& $1.51\,(0.05)$	& $1.68\,(0.06)$ \\
	& \textcolor{salmon}{\textbf{light red}}	& 6134	& 9.75	& 10.36	& 11.55	& -12.10	& -11.83	& -11.50	& $2.08\,(0.05)$	& $1.90\,(0.06)$ \\
	& \textcolor{red}{\textbf{red}}	& 4462	& 9.76	& 10.70	& 12.00	& -13.40	& -12.32	& -12.10	& $2.00\,(0.05)$	& $1.82\,(0.05)$ \\
$M_*/{\rm sSFR\ grid}$	& \textcolor{purple}{\textbf{purple}}	& 7599	& 9.25	& 9.47	& 9.75	& -10.00	& -9.74	& -8.60	& $0.65\,(0.03)$	& $1.04\,(0.05)$ \\
	& \textcolor{magenta}{\textbf{magenta}}	& 4401	& 9.25	& 9.51	& 9.75	& -10.80	& -10.28	& -10.00	& $1.17\,(0.03)$	& $1.51\,(0.05)$ \\
	& \textbf{black}	& 2846	& 9.75	& 9.98	& 10.40	& -10.00	& -9.82	& -8.41	& $0.73\,(0.03)$	& $1.05\,(0.05)$ \\
	& \textcolor{blue}{\textbf{blue}}	& 6668	& 9.75	& 10.06	& 10.40	& -10.80	& -10.33	& -10.00	& $0.79\,(0.03)$	& $1.23\,(0.05)$ \\
	& \textcolor{lime}{\textbf{light green}}	& 3940	& 9.75	& 10.06	& 10.40	& -11.70	& -11.29	& -10.80	& $1.87\,(0.05)$	& $1.89\,(0.06)$ \\
	& \textcolor{red}{\textbf{red}}	& 3104	& 9.75	& 10.15	& 10.40	& -13.40	& -11.97	& -11.70	& $2.82\,(0.07)$	& $2.11\,(0.06)$ \\
	& \textcolor{cyan}{\textbf{cyan}}	& 2262	& 10.40	& 10.61	& 11.35	& -10.80	& -10.44	& -10.00	& $0.86\,(0.03)$	& $1.21\,(0.05)$ \\
	& \textcolor{green}{\textbf{dark green}}	& 2037	& 10.40	& 10.68	& 11.43	& -11.70	& -11.26	& -10.80	& $1.15\,(0.04)$	& $1.47\,(0.06)$ \\
	& \textcolor{salmon}{\textbf{light red}}	& 5842	& 10.40	& 10.75	& 11.50	& -13.36	& -12.19	& -11.70	& $1.69\,(0.04)$	& $1.72\,(0.05)$ \\
\enddata
\tablenotetext{a}{One-halo bias is measured on scales of \onehalo and two-halo bias on scales of \twohalo.}
\end{deluxetable*}

\subsubsection{Mock Galaxy Samples}\label{subsec04:samples_mock}

We define galaxy samples in mock catalogs at $z=0$ to compare to our SDSS results, and at $z=0.45$ and $z=0.9$ to compare to C17's results for PRIMUS data at \zlow and \zhigh, respectively.
The mock galaxy sample cuts are given in Table~\ref{tab04:samples_mock}. These are the same as the cuts that subdivide the corresponding data samples, except for the cut that distinguishes star-forming from quiescent galaxies at each mock redshift (Equation~\ref{eq04:mock_ssfr_cut}). Our SDSS and especially C17's PRIMUS data samples span a range of redshifts and are classified as star-forming or quiescent based on Equation~\ref{eq04:sdss_ssfr_cut} (for SDSS data) and C17's Equation~1 (for PRIMUS data), both of which evolve linearly with redshift. As our mock catalogs are snapshots at single redshifts, Equation~\ref{eq04:mock_ssfr_cut} does not contain any redshift dependence beyond having a single version for each mock.

\begin{deluxetable*}{llrcrrrrrcr}[h]
\tablecaption{Mock galaxy samples.
\label{tab04:samples_mock}
}

\tablehead{
\colhead{\multirow{2}{*}{Run}} &
\colhead{\multirow{2}{*}{Sample}} &
\colhead{\multirow{2}{*}{$N_{\rm gal}$}} &
\multicolumn{3}{c}{\logm} &
\multicolumn{3}{c}{$\log($sSFR/yr$^{-1})$} &
\colhead{Satellite} &
\colhead{\multirow{2}{*}{Bias\tablenotemark{a}}} \\
\colhead{} & \colhead{} & \colhead{} &
\colhead{min} & \colhead{mean} & \colhead{max} &
\colhead{min} & \colhead{mean} & \colhead{max} &
\colhead{fraction} & \colhead{} \\
\cline{1-11}
\multicolumn{11}{c}{\vspace{-5pt}} \\
\multicolumn{11}{c}{$z=0$}
}

\startdata
${\rm SF/Q\ split}$	& \textcolor{blue}{\textbf{blue}}	& 49946	& 9.75	& 10.08	& 10.50	& -11.07	& -10.21	& -8.84	& $0.24$	& $1.21\,(0.07)$ \\
	& \textcolor{red}{\textbf{red}}	& 38107	& 9.75	& 10.16	& 10.50	& -12.89	& -11.68	& -10.73	& $0.46$	& $1.64\,(0.07)$ \\
${\rm Main\ sequence\ split}$	& \textcolor{darkblue}{\textbf{dark blue}}	& 55321	& 9.25	& 9.88	& 11.40	& -11.16	& -9.87	& -8.72	& $0.26$	& $1.25\,(0.05)$ \\
	& \textcolor{dodgerblue}{\textbf{light blue}}	& 55343	& 9.25	& 9.89	& 11.42	& -11.41	& -10.36	& -9.78	& $0.24$	& $1.19\,(0.07)$ \\
	& \textcolor{salmon}{\textbf{light red}}	& 31422	& 9.75	& 10.41	& 11.65	& -12.62	& -11.54	& -10.73	& $0.39$	& $1.59\,(0.04)$ \\
	& \textcolor{red}{\textbf{red}}	& 31439	& 9.75	& 10.41	& 11.65	& -13.24	& -12.11	& -11.36	& $0.50$	& $1.76\,(0.09)$ \\
${\rm sSFR\ cuts}$	& \textbf{black}	& 46367	& 9.25	& 9.65	& 11.15	& -10.00	& -9.77	& -9.0	& $0.28$	& $1.26\,(0.04)$ \\
	& \textcolor{blue}{\textbf{blue}}	& 38815	& 9.25	& 9.92	& 11.26	& -10.40	& -10.19	& -10.0	& $0.23$	& $1.18\,(0.07)$ \\
	& \textcolor{dodgerblue}{\textbf{light blue}}	& 21041	& 9.25	& 10.15	& 11.37	& -10.80	& -10.57	& -10.4	& $0.21$	& $1.19\,(0.08)$ \\
	& \textcolor{lime}{\textbf{light green}}	& 19976	& 9.75	& 10.28	& 11.54	& -11.50	& -11.17	& -10.8	& $0.34$	& $1.44\,(0.03)$ \\
	& \textcolor{salmon}{\textbf{light red}}	& 30157	& 9.75	& 10.37	& 11.46	& -12.10	& -11.83	& -11.5	& $0.44$	& $1.67\,(0.05)$ \\
	& \textcolor{red}{\textbf{red}}	& 17726	& 9.75	& 10.67	& 11.99	& -13.24	& -12.32	& -12.1	& $0.49$	& $1.82\,(0.12)$ \\
$\mstar/{\rm sSFR\ grid}$	& \textcolor{purple}{\textbf{purple}}	& 31878	& 9.25	& 9.48	& 9.75	& -10.00	& -9.74	& -8.72	& $0.29$	& $1.27\,(0.04)$ \\
	& \textcolor{magenta}{\textbf{magenta}}	& 18243	& 9.25	& 9.53	& 9.75	& -10.80	& -10.24	& -10.0	& $0.25$	& $1.14\,(0.07)$ \\
	& \textbf{black}	& 13296	& 9.75	& 9.98	& 10.40	& -10.00	& -9.82	& -8.84	& $0.25$	& $1.22\,(0.03)$ \\
	& \textcolor{blue}{\textbf{blue}}	& 30953	& 9.75	& 10.06	& 10.40	& -10.80	& -10.33	& -10.0	& $0.23$	& $1.19\,(0.08)$ \\
	& \textcolor{lime}{\textbf{light green}}	& 18301	& 9.75	& 10.07	& 10.40	& -11.70	& -11.30	& -10.8	& $0.41$	& $1.52\,(0.03)$ \\
	& \textcolor{red}{\textbf{red}}	& 15265	& 9.75	& 10.16	& 10.40	& -12.89	& -11.98	& -11.7	& $0.51$	& $1.72\,(0.11)$ \\
	& \textcolor{cyan}{\textbf{cyan}}	& 10660	& 10.4	& 10.62	& 11.37	& -10.80	& -10.44	& -10.0	& $0.19$	& $1.22\,(0.09)$ \\
	& \textcolor{green}{\textbf{dark green}}	& 9714	& 10.4	& 10.68	& 11.47	& -11.70	& -11.27	& -10.8	& $0.28$	& $1.41\,(0.03)$ \\
	& \textcolor{salmon}{\textbf{light red}}	& 24528	& 10.4	& 10.74	& 11.50	& -13.24	& -12.16	& -11.7	& $0.45$	& $1.79\,(0.08)$ \\
\cutinhead{$z=0.45$}
${\rm SF/Q\ split}$	& \textcolor{blue}{\textbf{blue}}	& 117645	& 10.5	& 10.71	& 11.00	& -10.82	& -10.01	& -8.57	& $0.21$	& $1.20\,(0.02)$ \\
	& \textcolor{red}{\textbf{red}}	& 116897	& 10.5	& 10.72	& 11.00	& -12.53	& -11.68	& -10.74	& $0.38$	& $1.65\,(0.05)$ \\
${\rm Main\ sequence\ split}$	& \textcolor{darkblue}{\textbf{dark blue}}	& 484447	& 8.5	& 9.6	& 10.50	& -9.85	& -9.19	& -8.1	& $0.28$	& $1.12\,(0.02)$ \\
	& \textcolor{dodgerblue}{\textbf{light blue}}	& 484462	& 8.5	& 9.61	& 10.50	& -10.70	& -9.66	& -8.78	& $0.34$	& $1.20\,(0.02)$ \\
	& \textcolor{salmon}{\textbf{light red}}	& 116975	& 10.1	& 10.64	& 11.60	& -11.84	& -11.36	& -10.62	& $0.33$	& $1.54\,(0.05)$ \\
	& \textcolor{red}{\textbf{red}}	& 116983	& 10.1	& 10.64	& 11.60	& -12.53	& -11.88	& -11.42	& $0.44$	& $1.75\,(0.10)$ \\
${\rm sSFR\ cuts}$	& \textbf{black}	& 119453	& 8.5	& 9.17	& 10.50	& -9.00	& -8.81	& -8.1	& $0.27$	& $1.04\,(0.02)$ \\
	& \textcolor{blue}{\textbf{blue}}	& 541553	& 8.5	& 9.53	& 10.50	& -9.60	& -9.32	& -9.0	& $0.32$	& $1.19\,(0.01)$ \\
	& \textcolor{dodgerblue}{\textbf{light blue}}	& 307452	& 8.57	& 9.91	& 10.50	& -10.60	& -9.84	& -9.6	& $0.32$	& $1.21\,(0.02)$ \\
	& \textcolor{lime}{\textbf{light green}}	& 38552	& 10.0	& 10.55	& 11.50	& -11.20	& -10.98	& -10.6	& $0.31$	& $1.45\,(0.03)$ \\
	& \textcolor{salmon}{\textbf{light red}}	& 139955	& 10.0	& 10.56	& 11.50	& -11.80	& -11.52	& -11.2	& $0.38$	& $1.58\,(0.06)$ \\
	& \textcolor{red}{\textbf{red}}	& 75249	& 10.0	& 10.77	& 11.50	& -12.53	& -12.00	& -11.8	& $0.44$	& $1.79\,(0.09)$ \\
$\mstar/{\rm sSFR\ grid}$	& \textcolor{purple}{\textbf{purple}}	& 194019	& 8.5	& 9.1	& 9.50	& -9.20	& -8.96	& -8.2	& $0.31$	& $1.10\,(0.01)$ \\
	& \textcolor{magenta}{\textbf{magenta}}	& 226615	& 8.53	& 9.26	& 9.50	& -10.20	& -9.46	& -9.2	& $0.36$	& $1.19\,(0.03)$ \\
	& \textbf{black}	& 69657	& 9.5	& 9.78	& 10.50	& -9.20	& -9.04	& -8.2	& $0.24$	& $1.07\,(0.03)$ \\
	& \textcolor{blue}{\textbf{blue}}	& 463458	& 9.5	& 9.95	& 10.50	& -10.20	& -9.63	& -9.2	& $0.30$	& $1.18\,(0.02)$ \\
	& \textcolor{lime}{\textbf{light green}}	& 45038	& 9.5	& 10.1	& 10.50	& -11.20	& -10.82	& -10.2	& $0.42$	& $1.42\,(0.05)$ \\
	& \textcolor{red}{\textbf{red}}	& 95546	& 9.5	& 10.21	& 10.50	& -12.20	& -11.54	& -11.2	& $0.47$	& $1.57\,(0.09)$ \\
	& \textcolor{cyan}{\textbf{cyan}}	& 90503	& 10.5	& 10.72	& 11.50	& -10.20	& -9.88	& -9.2	& $0.20$	& $1.17\,(0.02)$ \\
	& \textcolor{green}{\textbf{dark green}}	& 57092	& 10.5	& 10.84	& 11.50	& -11.20	& -10.56	& -10.2	& $0.23$	& $1.35\,(0.03)$ \\
	& \textcolor{salmon}{\textbf{light red}}	& 127046	& 10.5	& 10.82	& 11.50	& -12.20	& -11.71	& -11.2	& $0.36$	& $1.67\,(0.07)$ \\
\cutinhead{$z=0.9$}
${\rm SF/Q\ split}$	& \textcolor{blue}{\textbf{blue}}	& 44044	& 10.5	& 10.72	& 11.00	& -10.46	& -9.64	& -8.13	& $0.24$	& $1.48\,(0.03)$ \\
	& \textcolor{red}{\textbf{red}}	& 62842	& 10.5	& 10.76	& 11.00	& -11.90	& -11.29	& -10.38	& $0.27$	& $1.82\,(0.06)$ \\
${\rm Main\ sequence\ split}$	& \textcolor{darkblue}{\textbf{dark blue}}	& 106422	& 8.88	& 9.87	& 10.50	& -9.49	& -8.91	& -7.98	& $0.27$	& $1.30\,(0.04)$ \\
	& \textcolor{dodgerblue}{\textbf{light blue}}	& 106430	& 8.88	& 9.88	& 10.50	& -10.36	& -9.31	& -8.25	& $0.34$	& $1.38\,(0.05)$ \\
	& \textcolor{salmon}{\textbf{light red}}	& 48400	& 10.1	& 10.81	& 11.60	& -11.59	& -11.07	& -10.36	& $0.23$	& $1.80\,(0.08)$ \\
	& \textcolor{red}{\textbf{red}}	& 48410	& 10.1	& 10.81	& 11.60	& -11.90	& -11.51	& -10.82	& $0.28$	& $1.91\,(0.08)$ \\
${\rm sSFR\ cuts}$	& \textbf{black}	& 56452	& 9.0	& 9.58	& 11.00	& -8.90	& -8.66	& -8.0	& $0.24$	& $1.16\,(0.04)$ \\
	& \textcolor{blue}{\textbf{blue}}	& 158677	& 9.01	& 10.04	& 11.00	& -9.60	& -9.25	& -8.9	& $0.32$	& $1.40\,(0.04)$ \\
	& \textcolor{dodgerblue}{\textbf{light blue}}	& 39763	& 9.51	& 10.55	& 11.00	& -10.20	& -9.78	& -9.6	& $0.29$	& $1.45\,(0.06)$ \\
	& \textcolor{lime}{\textbf{light green}}	& 8828	& 10.2	& 10.72	& 11.69	& -10.80	& -10.58	& -10.2	& $0.25$	& $1.63\,(0.07)$ \\
	& \textcolor{salmon}{\textbf{light red}}	& 26630	& 10.2	& 10.72	& 11.65	& -11.20	& -11.03	& -10.8	& $0.24$	& $1.75\,(0.07)$ \\
	& \textcolor{red}{\textbf{red}}	& 60417	& 10.2	& 10.86	& 11.69	& -11.80	& -11.46	& -11.2	& $0.26$	& $1.92\,(0.09)$ \\
$\mstar/{\rm sSFR\ grid}$	& \textcolor{purple}{\textbf{purple}}	& 29931	& 8.88	& 9.33	& 9.50	& -9.20	& -8.67	& -8.2	& $0.26$	& $1.12\,(0.03)$ \\
	& \textbf{black}	& 89738	& 9.5	& 9.84	& 10.50	& -9.20	& -8.95	& -8.2	& $0.30$	& $1.34\,(0.06)$ \\
	& \textcolor{blue}{\textbf{blue}}	& 90797	& 9.5	& 10.1	& 10.50	& -10.20	& -9.44	& -9.2	& $0.34$	& $1.39\,(0.03)$ \\
	& \textcolor{lime}{\textbf{light green}}	& 9133	& 9.73	& 10.33	& 10.50	& -11.20	& -10.86	& -10.2	& $0.33$	& $1.58\,(0.05)$ \\
	& \textcolor{cyan}{\textbf{cyan}}	& 48638	& 10.5	& 10.79	& 11.50	& -10.20	& -9.69	& -9.2	& $0.23$	& $1.50\,(0.03)$ \\
	& \textcolor{green}{\textbf{dark green}}	& 27713	& 10.5	& 10.81	& 11.50	& -11.20	& -10.93	& -10.2	& $0.23$	& $1.76\,(0.04)$ \\
	& \textcolor{salmon}{\textbf{light red}}	& 59321	& 10.5	& 10.89	& 11.50	& -11.90	& -11.48	& -11.2	& $0.26$	& $1.93\,(0.10)$ \\
\enddata
\tablenotetext{a}{These measurements are made on scales of \twohalo.}
\end{deluxetable*}

\section{Methods}\label{sec04:methods}

In this section we describe the methods used to measure the projected correlation functions and bias values of galaxy samples in both SDSS data and mock galaxy catalogs. We also describe how we estimate the errors of these measurements, including uncertainties due to cosmic variance. 

\subsection{Clustering Measurements}\label{subsec04:wp}

To measure projected two-point clustering in SDSS, \wprp, we use the {\tt correl} program in \UM. The program uses the \citet{landy_szalay93} estimator ${(DD - 2DR+RR)}$ to compute the redshift-space correlation function, \xirppi, which is then integrated over ${\vert\pi\vert < 20}$ \Mpch to compute the projected correlation function \wprp. The code uses $10^6$ random points drawn from the same mask region with uniform volume distribution to compute $DR$, while $RR$ is computed via Monte Carlo integration. Jackknife resampling is used to estimate errors, giving us an estimate for the lower bound of samples with volumes $V_{\rm eff}=0.3 {\rm Gpc}^3.$

In the mock catalogs, to reduce the Poisson errors we estimate the two-point correlation function \xir of mock galaxy samples by measuring the autocorrelation function (ACF) of all mock galaxies, and the cross-correlation function (CCF) between each sample and all galaxies in the mock. We then infer the ACF of each sample as described below. 

For ACF and CCF measurements we use the \halotools \citep{hearin_etal17} function \texttt{wp\_jackknife} with the  \citet{davis_peebles83} estimator: $\xir = DD(r)/DR(r) - 1$. For ACF measurements $DD(r)$ counts pair separations among all galaxies in a given sample, while for CCF measurements $DD(r)$ is a count of pair separations between the galaxy sample of interest and a ``tracer" galaxy sample consisting of the entire mock catalog.

We measure \xir separately both perpendicular to (\rp) and along ($\pi$) the mock catalog's line-of-sight dimension, then integrate \xirppi over the line-of-sight to a given value of \pimax to obtain the projected correlation function \wprp. For each mock we use the same \pimax value used for \wprp measurements in the corresponding dataset:\ $\pimax=20\ \Mpch$ for the $z=0$ mock and SDSS data,\footnote{For SDSS data we tested both $\pimax=20$ and $40\ \Mpch$ and found that the signal-to-noise of individual \wprp measurements is almost universally larger for $\pimax=20\ \Mpch$.} and $\pimax=40\ \Mpch$ for the $z=0.45$ and $z=0.9$ mocks. The latter is consistent with C17 and their clustering measurements of PRIMUS galaxy samples.

The inferred projected ACF \wprp for each mock galaxy sample is
\begin{equation}
    \wprp = \frac{\omega^2_{\rm GT}(\rp)}{\omega_{\rm TT}(\rp)},
\end{equation}
\noindent where $\omega_{\rm GT}(\rp)$ is the projected galaxy--tracer CCF, and $\omega_{\rm TT}(\rp)$ is the projected tracer--tracer ACF. Inferring the ACF in this way reduces the error on \wprp, especially for smaller galaxy samples.

The \texttt{wp\_jackknife} function estimates the error of \wprp by dividing the mock volume $N$ times along each dimension to define $N_j=(N+1)^3$ equal subvolumes and creates the same number of jackknife samples, where each jackknife sample is the entire mock volume excluding one subvolume.
The error of \wprp is then $[\sigma_{\wp}^2 (N_j-1)/N_j ]^{1/2}$, where $\sigma_{\wp}^2$ is the variance of \wprp across the jackknife samples.

\subsection{Absolute and Relative Bias Measurements}\label{subsec04:bias}

We measure the absolute bias of both SDSS and mock galaxy samples using the projected correlation function \wprp of each sample. Absolute bias is a measure of the clustering strength of a particular galaxy sample compared to that of dark matter, and is defined as $\sqrt{\omega_{\rm G}/\omega_{\rm DM}}$, where $\omega_{\rm G}$ and $\omega_{\rm DM}$ are the galaxy and dark matter projected correlation functions, respectively, averaged over ``two-halo" scales of \twohalo.
To estimate $\omega_{\rm DM}$ we use the publicly available code of \citet{smith_etal03} to calculate $\xi_{\rm DM}$ at the median redshift of the relevant galaxy sample, then integrate $\xi_{\rm DM}$ to the same value of \pimax used for the corresponding galaxy sample.

The relative bias of two galaxy samples is the square root of the ratio of their respective projected correlation functions, averaged over a given length scale, and compares the clustering strengths of the two samples on that scale. We measure the relative bias as a function of scale between pairs of SDSS galaxy samples and pairs of mock galaxy samples, $b_{\rm rel}(\rp)$, as
\begin{equation}\label{eq:b_rel}
b_{\rm rel}(\rp) = \sqrt{\frac{\omega_1(\rp)}{\omega_2(\rp)}},
\end{equation}
\noindent where $\omega_1(\rp)$ and $\omega_2(\rp)$ are the projected correlation functions of the two samples. We then take the average of Equation~\ref{eq:b_rel} over \onehalo at a given redshift to estimate the ``one-halo" relative bias, and average over \twohalo to estimate the ``two-halo" term of the relative bias.

To estimate the error on absolute and relative bias measurements of SDSS galaxy samples we calculate the bias $b$ for \emph{each} jackknife sample and compute the variance of the relevant bias itself across all the jackknife samples. The error of that bias measurement is then $\sqrt{{(N_j-1}/N_j)\,\sigma_b^2}$, where $N_j$ is again the total number of jackknife samples, and $\sigma_b^2$ is the variance of $b$ across the samples.
This method does not use the error on \wprp described in the previous section, and instead estimates the uncertainty on the relative bias directly, accounting for cosmic variance across jackknife samples.

\section{Correlation Functions and Bias of Galaxy Samples in Data and Mocks}\label{sec04:wp_bias}

In this section we present measurements of \wprp, and of absolute and relative bias versus stellar mass and sSFR, for our SDSS galaxy samples. We also present the same measurements for mock galaxy catalogs at $z=0$, 0.45, and 0.9, and compare these results to the corresponding data at each redshift.

\subsection{SDSS Clustering Dependence on Stellar Mass and sSFR}\label{subsec04:results_sdss}

\begin{figure*}[t!]
\centering
\includegraphics[width=0.9\linewidth]{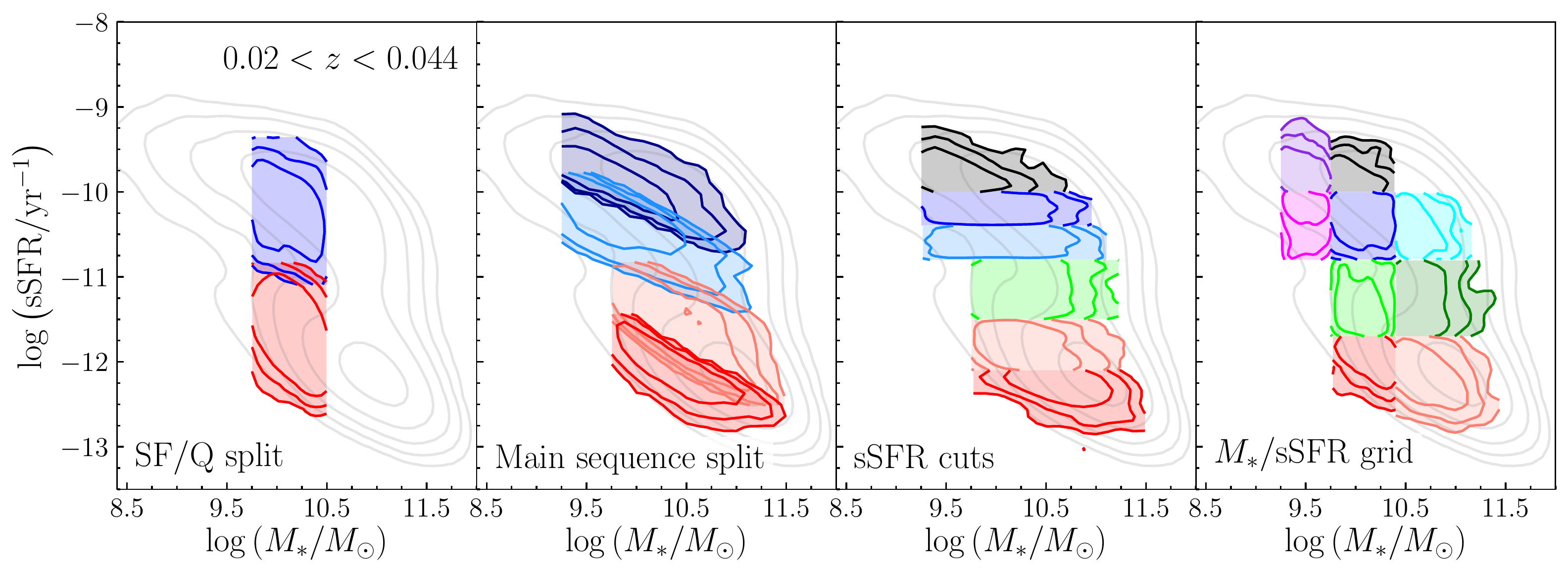}
\includegraphics[width=0.9\linewidth]{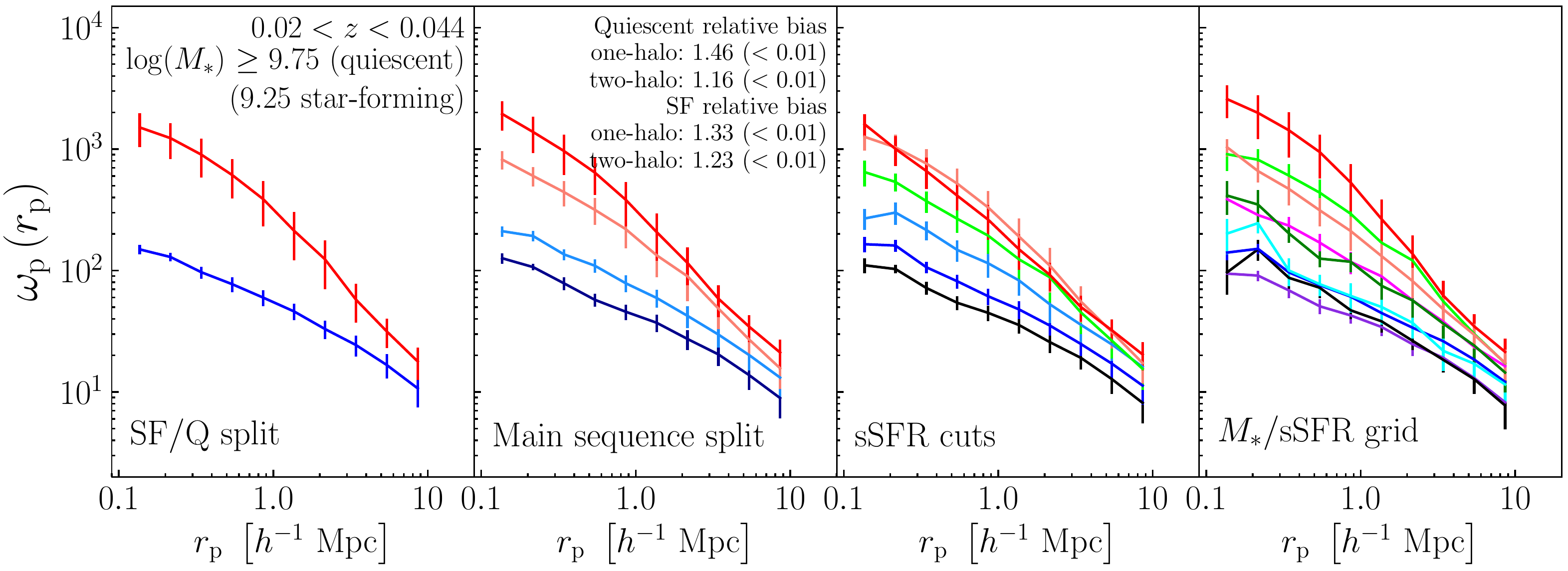}
\caption{
Top row:\ Specific star formation rate (sSFR) versus stellar mass for all SDSS galaxy samples used here. From left to right galaxies are divided into (i) star-forming versus quiescent (``SF/Q split"), (ii) above versus below the star-forming and quiescent main sequences (``main sequence split"), (iii) cuts in sSFR (``sSFR cuts"), and (iv) cuts in both sSFR and stellar mass (``\mstar/sSFR grid").
All samples are stellar mass complete, with ${\logm > 9.25}$ for star-forming and ${\logm > 9.75}$ for quiescent galaxy samples (see text for details).
Bottom row:\ The projected correlation function \wprp for each of the galaxy samples shown in the top row. The relative biases (see \S\ref{subsec04:bias}) on one-halo and two-halo scales of ``main sequence split" samples are given. Errors on \wprp are estimated by jackknife resampling.
}
\label{fig04:sdss_samples_sm9p75q_9p25sf}
\end{figure*}

Figure~\ref{fig04:sdss_samples_sm9p75q_9p25sf} shows the stellar mass and sSFR distributions of all SDSS galaxy samples, as well as the projected correlation function \wprp of each sample. Table~\ref{tab04:samples_sdss} provides the bias on one-halo (\onehalo) and two-halo (\twohalo) scales for each sample.

The star-forming/quiescent split samples clearly show that quiescent galaxies are more strongly clustered than star-forming galaxies at fixed stellar mass.
This confirms previous studies of the dependence of SDSS clustering on galaxy color \citep[e.g.][]{heinis_etal09, hearin_etal14, watson_etal15}, often used as a proxy for SFR.

The ``main sequence split" samples in Figure~\ref{fig04:sdss_samples_sm9p75q_9p25sf} (second column) show that clustering strength is correlated with sSFR at fixed stellar mass \emph{within} both the star-forming and quiescent populations.
Within the star-forming main sequence, galaxies below the sequence are substantially more strongly clustered than galaxies above the sequence. The relative bias between 
these samples is ${1.33\pm0.02}$ on one-halo scales and ${1.23\pm0.03}$ on two-halo scales.
We find similar results within the quiescent population, where galaxies below the quiescent sequence are more clustered than galaxies above the sequence, which have higher sSFR at a given stellar mass. 
This is the first time this has been shown in SDSS and reflects the trends seen at higher redshift in \citet{mostek_etal13} and C17.

The third set of samples (``sSFR cuts") display an anticorrelation between sSFR and clustering strength across the full galaxy population, although the trend is more pronounced for star-forming galaxies. In the lower panel of the third column of Figure~\ref{fig04:sdss_samples_sm9p75q_9p25sf} the amplitude of \wprp increases smoothly as sSFR decreases across five of the six samples, from the highest sSFR sample (dark blue) to the second lowest sSFR sample (light red). The lowest sSFR sample (red) is similar to that of the next lowest sSFR sample (light red); within the errors \wprp is the same for these two samples on both one-halo and two-halo scales. 
The lack of differentiation of \wprp for the two lowest sSFR samples could be due in part to the difficulty of estimating robust SFRs for galaxies with very low star formation rates.

The final set of samples (``stellar mass/sSFR grid") are most easily interpreted by considering separately subsets confined to either a given stellar mass or sSFR bin. 
Following C17, these samples are used primarily to fill out the range of stellar mass and sSFR \emph{ratios} over which we explore the dependence of clustering on sSFR at fixed stellar mass and on stellar mass at fixed sSFR in the following sections.

\begin{figure*}[t!]
\centering
\includegraphics[width=0.9\linewidth]{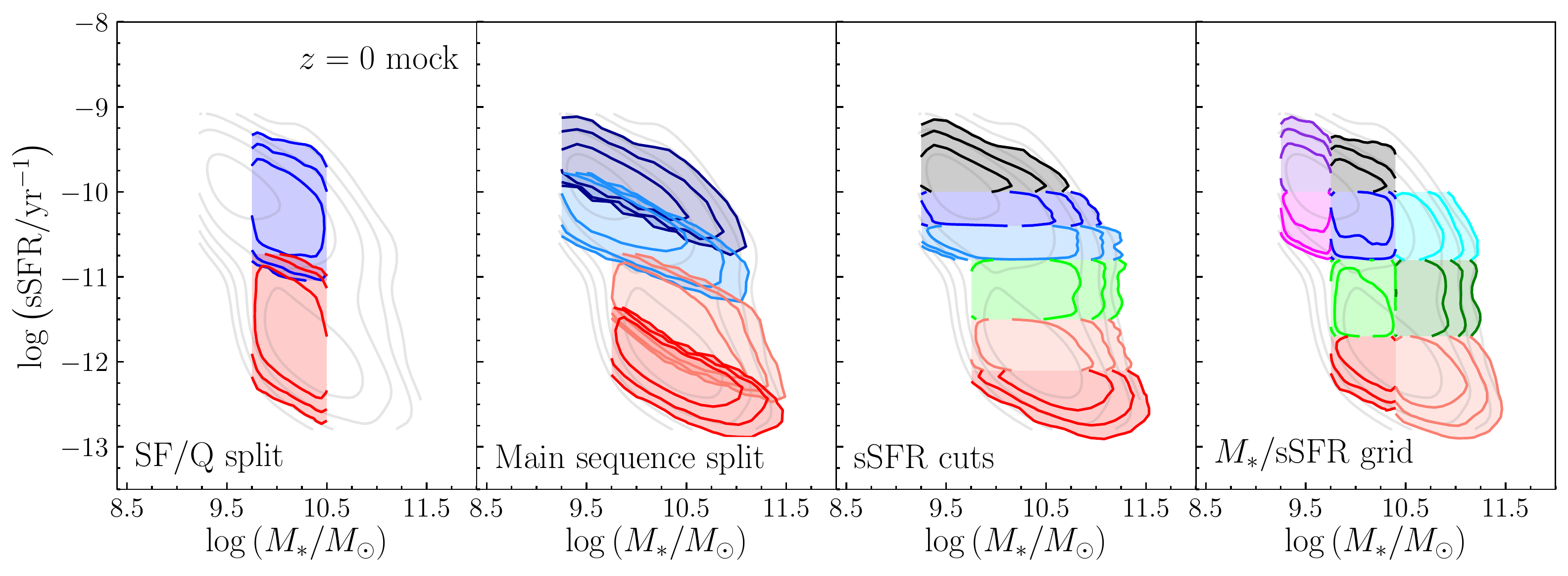}
\includegraphics[width=0.9\linewidth]{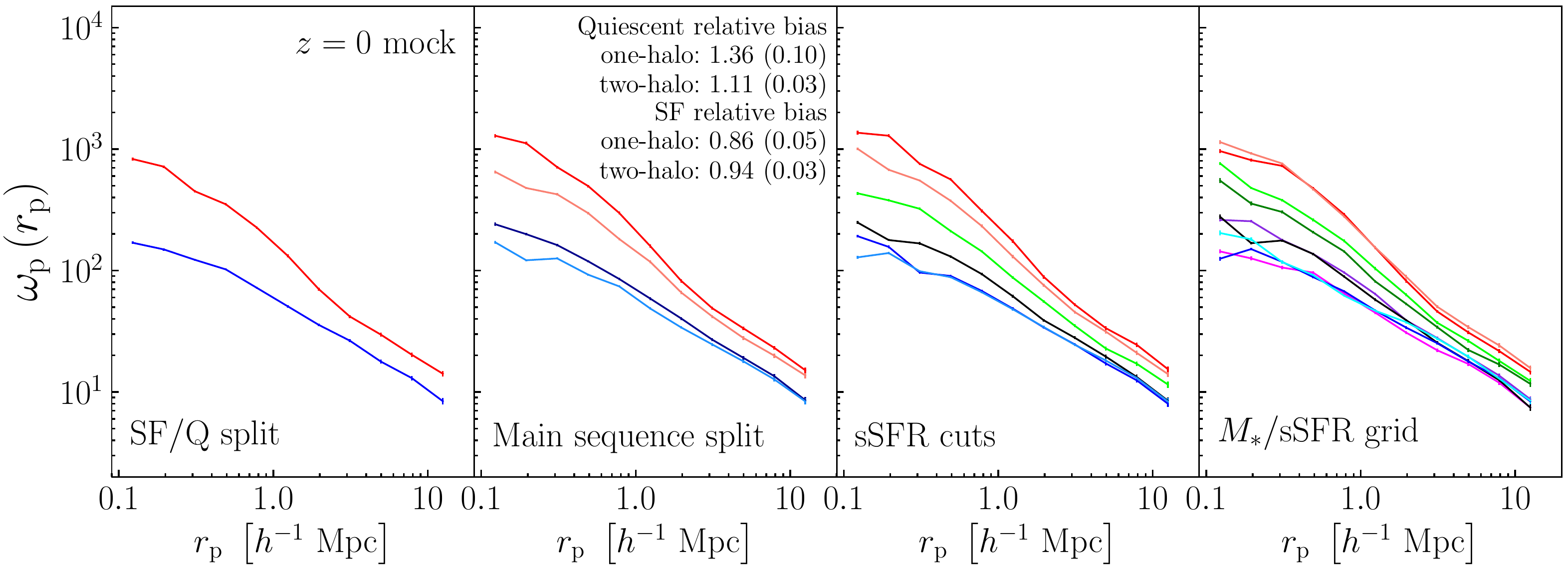}
\caption{
Analogous to Figure~\ref{fig04:sdss_samples_sm9p75q_9p25sf} but for the $z=0$ mock galaxy catalog.
The top row shows mock galaxy sample distributions in the stellar mass--sSFR plane. All samples are stellar mass complete, as described in the text.
The bottom row shows the projected correlation function \wprp for each sample show in the top row, and the relative biases (see \S\ref{subsec04:bias}) on one-halo and two-halo scales of ``main sequence split" samples are given in the second panel.
}
\label{fig04:samples_z0p0_obs}
\end{figure*}

\begin{figure*}[t!]
\centering
\includegraphics[width=0.9\linewidth]{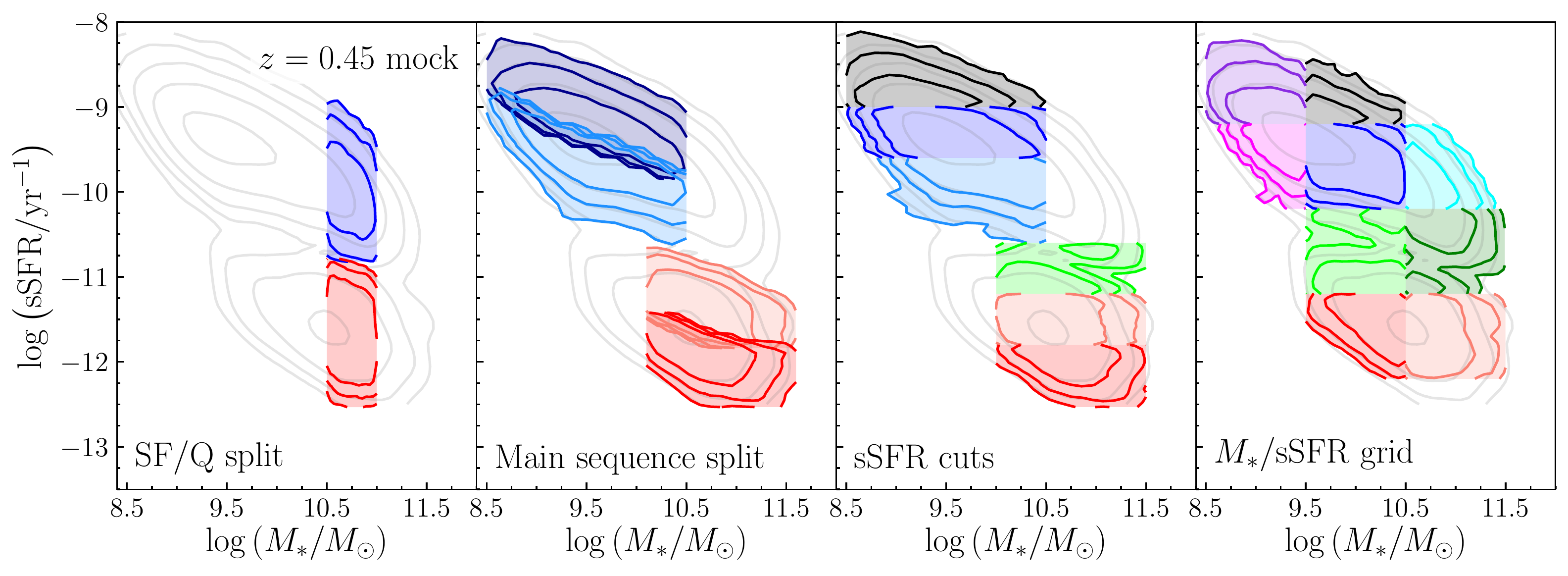}
\includegraphics[width=0.9\linewidth]{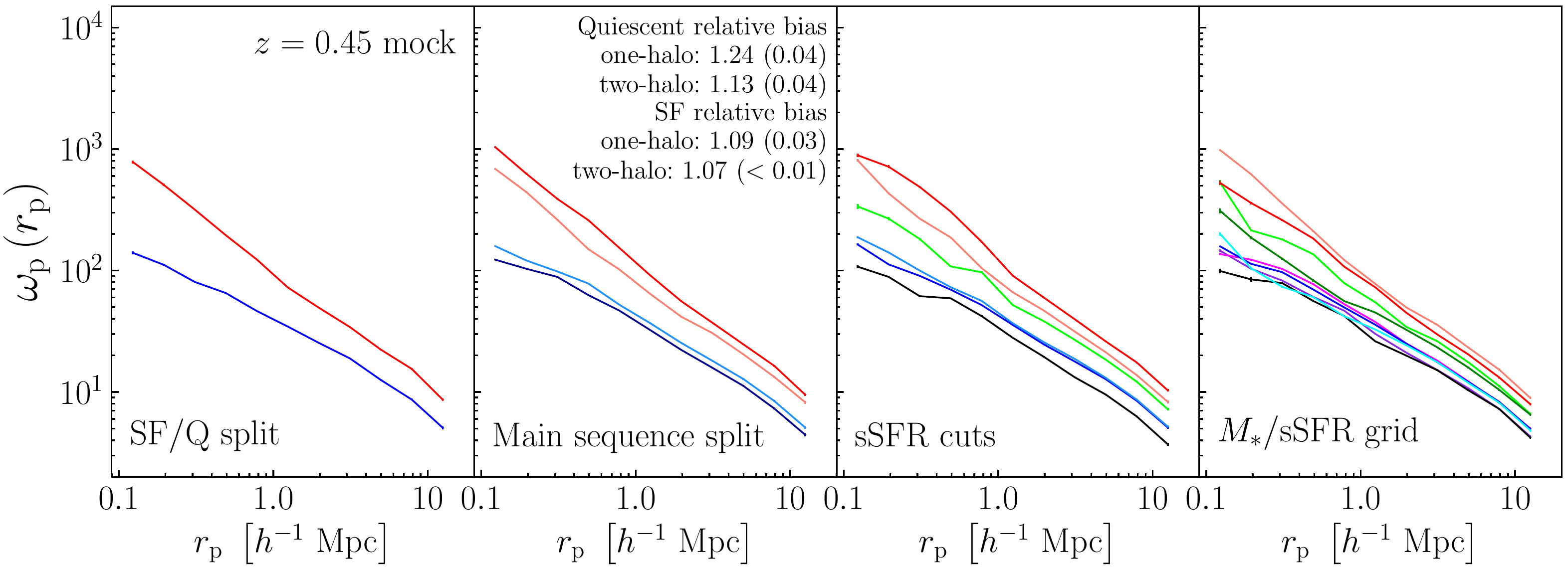}
\caption{
Same as Figure~\ref{fig04:samples_z0p0_obs} but for the $z=0.45$ mock galaxy catalog. Only the star-forming/quiescent split samples (first column) are stellar mass complete. We compare these results to Figure 2 of C17, which shows analogous measurements for PRIMUS galaxy samples at \zlow and \zhigh.
}
\label{fig04:samples_z0p45_obs}
\end{figure*}

\subsection{Projected Correlation Functions in Mocks}\label{subsec04:mocks_wp}

In this section we compare the clustering measurements in SDSS and PRIMUS data to equivalent measurements in mock galaxy catalogs at $z=0$, $z=0.45$, and $z=0.9$. 
Table~\ref{tab04:samples_mock} lists the details and two-halo bias measurements of all mock galaxy samples.

The \UM model is observationally constrained in part by measurements of stellar mass-complete clustering in SDSS data (at $z\sim0$) and PRIMUS data (at $z\sim0.45$). The model utilizes clustering measurements of star-forming and quiescent galaxies separately, but does \emph{not} directly incorporate measurements of clustering dependence in narrower bins in sSFR or the intra-sequence relative bias previously observed in PRIMUS. A natural question is whether and to what extent does \UM reproduce the variation in clustering strength with sSFR observed \emph{within} the star-forming and quiescent galaxy populations in both SDSS and PRIMUS data?

Figure~\ref{fig04:samples_z0p0_obs} is analogous to Figure~\ref{fig04:sdss_samples_sm9p75q_9p25sf} and shows the sSFR and stellar mass distributions and projected correlation functions of galaxy samples in the $z=0$ mock described above in \S\ref{sec04:data_sims}. There is excellent agreement between SDSS and the $z=0$ mock for the star-forming/quiescent split galaxy samples, largely by design. 
The amplitude of the \wprp one-halo term for quiescent mock galaxies is smaller than for the analogous SDSS galaxy sample. However, it is known that the clustering amplitude of SDSS galaxies is higher at $z<0.05$ (e.g.\ Figure C5 of \citet{behroozi_etal19} and the associated discussion).

Clustering in the SDSS data and the $z=0$ mock further diverge when we consider the ``main sequence split" samples. The relative bias within the star-forming $z=0$ mock galaxy population---i.e.\ the bias ratio of galaxies with above average to below average sSFRs---is less than unity on both one-halo and two-halo scales, while in the data it is greater than unity. We discuss this reversal below in \S\ref{sec04:z0_mod}.

Two differences between the $z=0$ mock and SDSS data are present in the third set of samples (``sSFR cuts"), which divide the galaxy population into six bins in sSFR. In the mock the clustering strength of the three samples spanning the quiescent sequence correlates with sSFR, especially on one-halo scales. The lowest sSFR quiescent sample (red) has the largest \wprp, followed by the next lowest sSFR sample (light red). The highest sSFR quiescent sample (light green) is the least strongly clustered of the three quiescent mock samples, and splits the difference in \wprp amplitude between the other quiescent samples and the three star-forming samples.
In contrast, the quiescent ``sSFR cuts" samples in SDSS data are less differentiated in terms of relative clustering strength than the corresponding mock samples. The projected correlation functions of lowest and second lowest sSFR samples agree within the errors, on both one-halo and two-halo scales. As previously discussed in \S\ref{subsec04:results_sdss}, the lack of differentiation in \wprp between the two lowest sSFR samples in SDSS may be at least partly be due to the difficulty of robustly inferring SFRs for SDSS galaxies with the lowest rates of star formation.

At higher redshift we compare our clustering results in mock catalogs to C17's clustering measurements of analogous galaxy samples in PRIMUS data at \zlow (which we compare to the $z=0.45$ mock) and at \zhigh (which we compare to the $z=0.9$ mock). Interestingly, there are different trends in terms of agreement between the higher redshift mocks and PRIMUS data than for the $z=0$ mock and SDSS data.

As expected, the star-forming/quiescent split mock samples at both $z=0.45$ (Figure~\ref{fig04:samples_z0p45_obs}) and $z=0.9$ agree with what C17 finds in PRIMUS data at \zlow and \zhigh:\ quiescent galaxies are more strongly clustered than star-forming galaxies at fixed stellar mass. This agreement is unsurprising, as the star-forming/quiescent split sample results at $z\sim0.45$ from C17 serve as constraints for the \UM model.

The ``main sequence split" mock samples divide the star-forming and quiescent populations each into two samples with identical stellar mass distributions, allowing us to compare the clustering of star-forming galaxies with above average and below average sSFRs, independent of stellar mass, and likewise for quiescent galaxies. We find a correlation between clustering amplitude and sSFR \emph{within} both the star-forming and quiescent sequences at $z=0.45$ and $z=0.9$, although the magnitude of the effect is stronger at $z=0.45$, particularly for the quiescent population. These results again agree with what C17 find in PRIMUS data:\ on both one-halo and two-halo scales star-forming galaxies with the highest sSFRs are less clustered than those with sSFRs below the SFMS. Similarly, quiescent galaxies with the lowest sSFRs are more clustered than those with above average sSFRs.

The third set of galaxy samples (``sSFR cuts") also display qualitative agreement between the mocks and the corresponding PRIMUS data samples at both \zlow and \zhigh, although direct comparisons of \wprp for the \zhigh data samples and $z=0.9$ are complicated by the higher noise in the data samples. At \zlow C17 find a general decline in the amplitude of \wprp with increasing sSFR, although the two lowest sSFR samples have the same one-halo clustering amplitude within the errors, and two of the three star-forming samples have nearly identical clustering strengths on two-halo scales. In the $z=0.45$ mock the corresponding two star-forming samples also have nearly identical clustering amplitudes on both one-halo and two-halo scales. The highest sSFR star-forming mock sample has the lowest clustering amplitude, which C17 also find for PRIMUS data. While in the $z=0.45$ mock we see a clear decline in clustering strength with increasing sSFR across the three quiescent samples, this distinction is less prominent in the corresponding quiescent data samples.

At \zhigh the projected correlation functions of the PRIMUS ``sSFR cuts" samples are noisier, and a more useful comparison to the corresponding mock is made by comparing the absolute and relative biases, as performed in the following subsections. 

\subsection{Absolute Bias of Galaxy Samples in Data and Mocks}\label{subsec04:abs_bias}

\begin{figure*}[t]
\centering
\includegraphics[width=\linewidth]{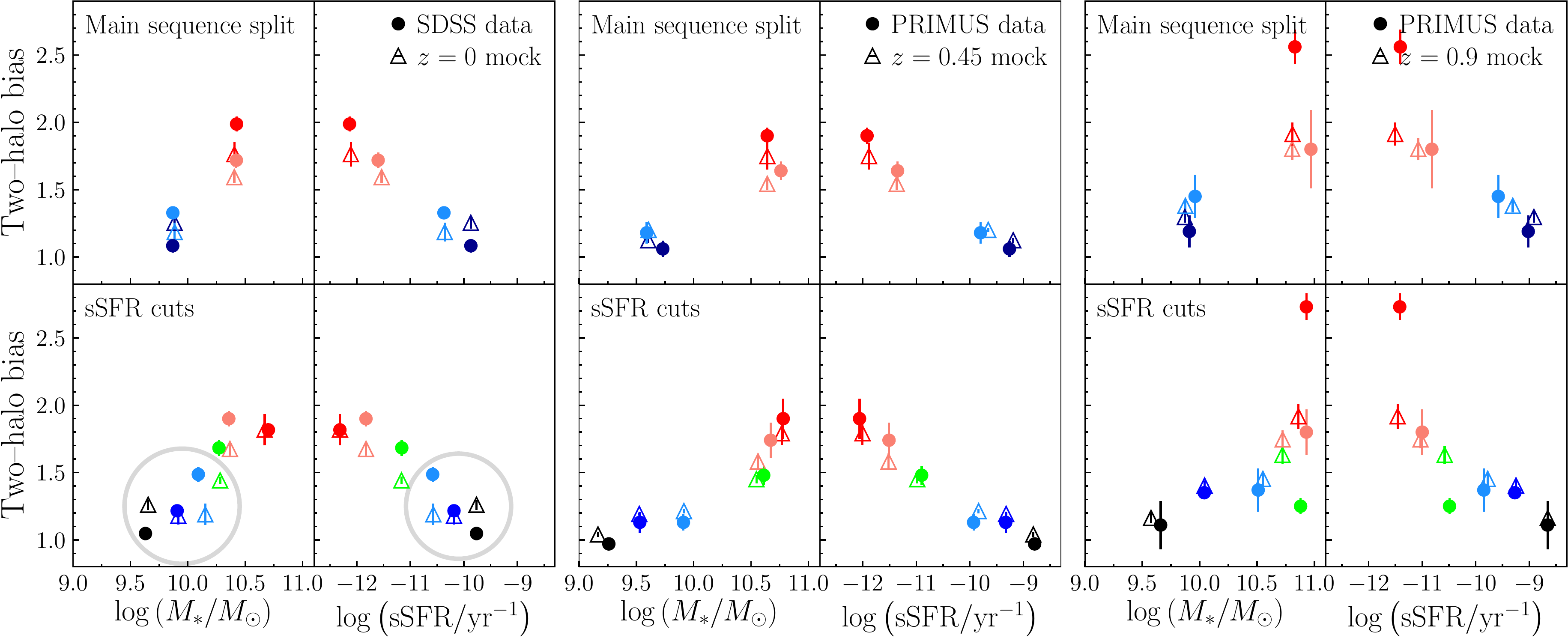}
\caption{
Absolute bias measured on two-halo scales as a function of mean stellar mass (left column in each set of panels) and mean sSFR (right column in each set of panels) for data (filled circles) and mock (open triangles) galaxy samples. Shown are the ``main sequence split" (top row) and ``sSFR cuts" (bottom row) samples for $z=0$ (left panels), $z=0.45$ (middle panels), and $z=0.9$ (right panels). The colors used are the same as in Figures~\ref{fig04:sdss_samples_sm9p75q_9p25sf}, \ref{fig04:samples_z0p0_obs}, and \ref{fig04:samples_z0p45_obs}.
Errors are estimated by jackknife resampling. Gray circles highlight the area of disagreement between the data and model at $z=0$, discussed further in the text.
}
\label{fig04:abs_bias_data_vs_mocks}
\end{figure*}

Figure~\ref{fig04:abs_bias_data_vs_mocks} shows the absolute bias on two-halo scales of the ``main sequence split" and ``sSFR cuts" galaxy samples for both data and mocks at $z\sim0$, $z\sim0.45$, and $z\sim0.9$.
At $z\sim0$ there is an overall normalization difference between the data and mock that is not present at higher redshift:\ the bias values in the mock samples are generally lower than in the corresponding data sample. However it is known that SDSS data exhibits a clustering excess at $z<0.05$ \citep[see Figure C5 of][]{behroozi_etal19}, which aligns with the redshift range used here, and is consistent with the bias offsets between the data and $z=0$ mock galaxy samples shown in Figure~\ref{fig04:abs_bias_data_vs_mocks}.

As an additional check we also tested using time-averaged (versus instantaneous) SFRs for the $z=0$ mock. While this did slightly improve agreement with the data, the overall trends were the same as using those found using instantaneous SFRs.

Up to the overall normalization difference discussed above, the data and $z=0$ mock agree well for quiescent galaxy samples (red, light red, and green points in Figure~\ref{fig04:abs_bias_data_vs_mocks}). The one exception is the two ``sSFR cuts" samples with the lowest sSFRs (red and light red), which are reversed from the general trend:\ the lowest sSFR sample is less biased than the next lowest sSFR sample. As discussed in \S\ref{subsec04:results_sdss} above, this may be due to the difficulty of robustly estimating SFRs for galaxies with the lowest rates of star formation with sufficient accuracy to meaningfully differentiate between the galaxies in these two samples.

For star-forming galaxy samples (light blue, blue, and dark blue points in Figure~\ref{fig04:abs_bias_data_vs_mocks}, highlighted by the two gray circles in the lower left panels) the data and $z=0$ mock do not agree. In the data the bias decreases monotonically with increasing sSFR, while in the mock the highest sSFR samples are \emph{more} biased than star-forming galaxies with lower sSFRs, i.e.\ the opposite trend. This discrepancy is not seen at the higher redshifts.

Agreement between the data and corresponding mock is strongest at $z\sim0.45$, where the data and mock values match within the errors for every galaxy sample.  
The $z=0.9$ mock samples also agree well with the corresponding data samples, particularly for star-forming galaxies. Within the quiescent population the lowest sSFR data sample is substantially more biased in the data than in the mock, both when quiescent galaxies are split into two samples above and below the quiescent main sequence in the stellar mass--sSFR plane, and into three samples using simple cuts in sSFR. As noted above, at these higher redshifts the PRIMUS results are noisier, due to smaller sample sizes.

\subsection{Relative Bias of Galaxy Samples in Data and Mocks}\label{subsec04:rel_bias}

\begin{figure*}[t]
\normalsize
\centering
\setlength{\tabcolsep}{0pt}
\begin{tabular}{p{0.4275\textwidth} p{0.0225\textwidth} p{0.4275\textwidth}}
\includegraphics[width=\linewidth]{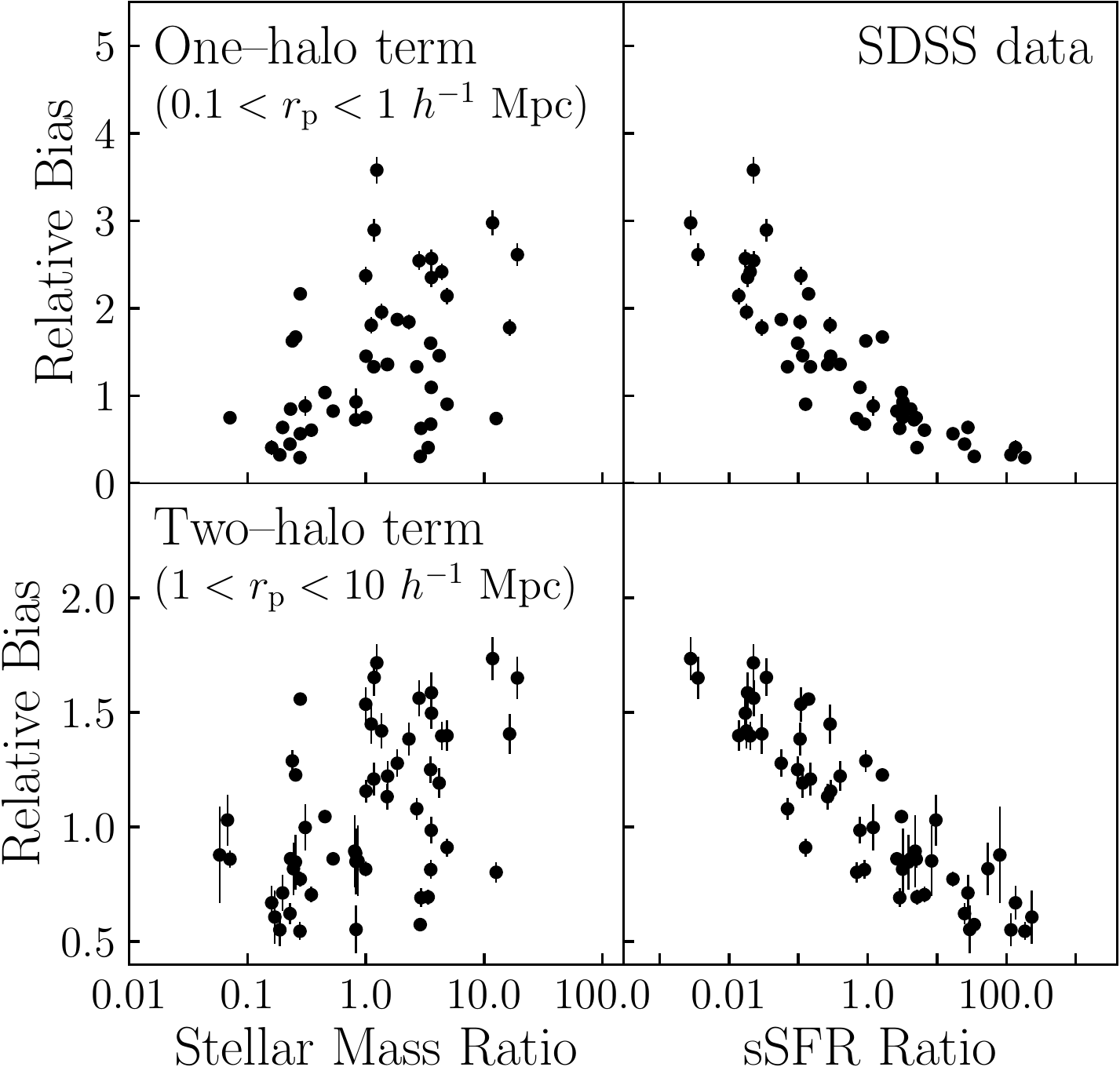} & &
\includegraphics[width=\linewidth]{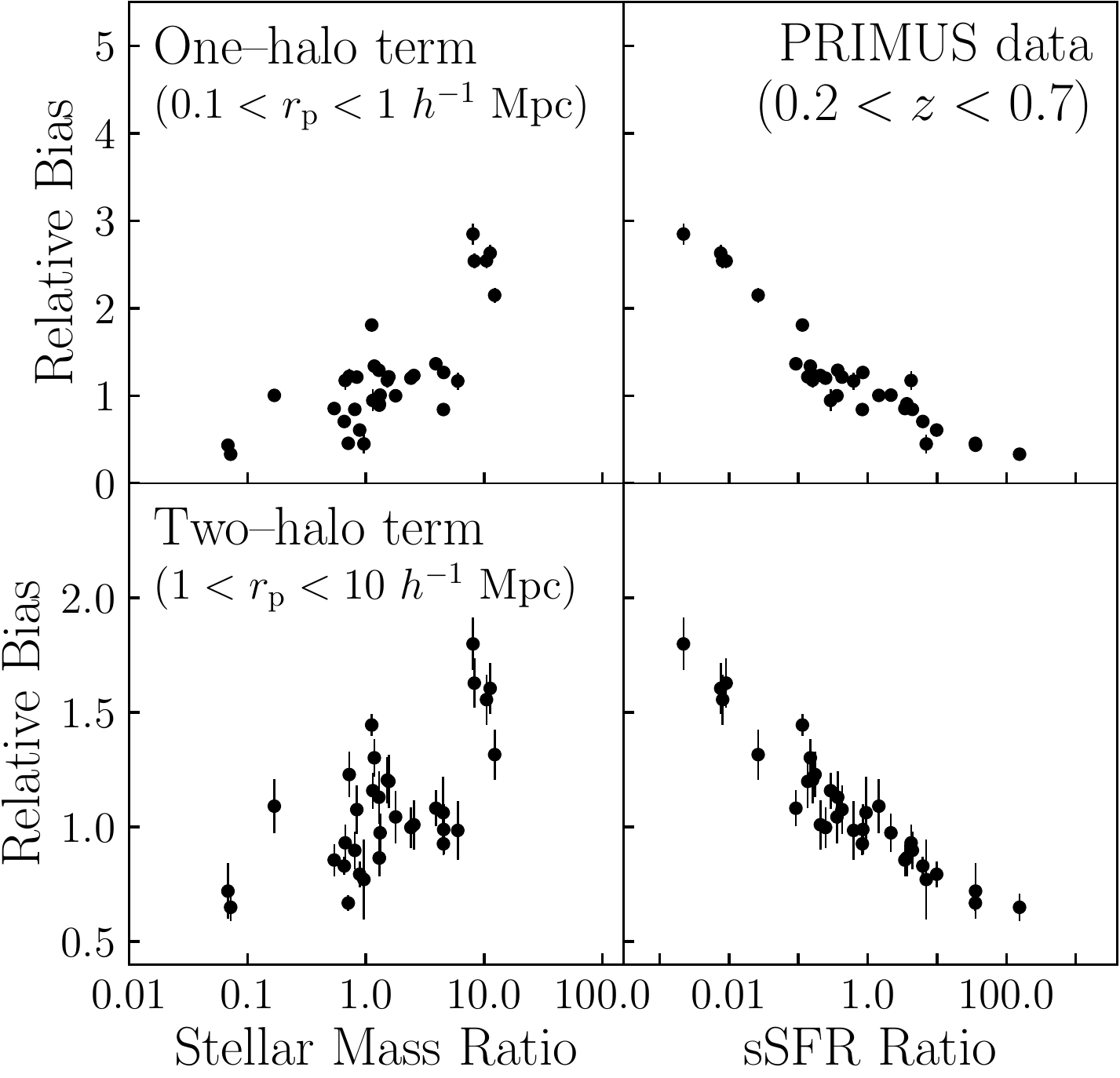} \\
\multicolumn{1}{c}{\textbf{(a)} SDSS data} & &
\multicolumn{1}{c}{\textbf{(b)} PRIMUS data} \\ \\
\includegraphics[width=\linewidth]{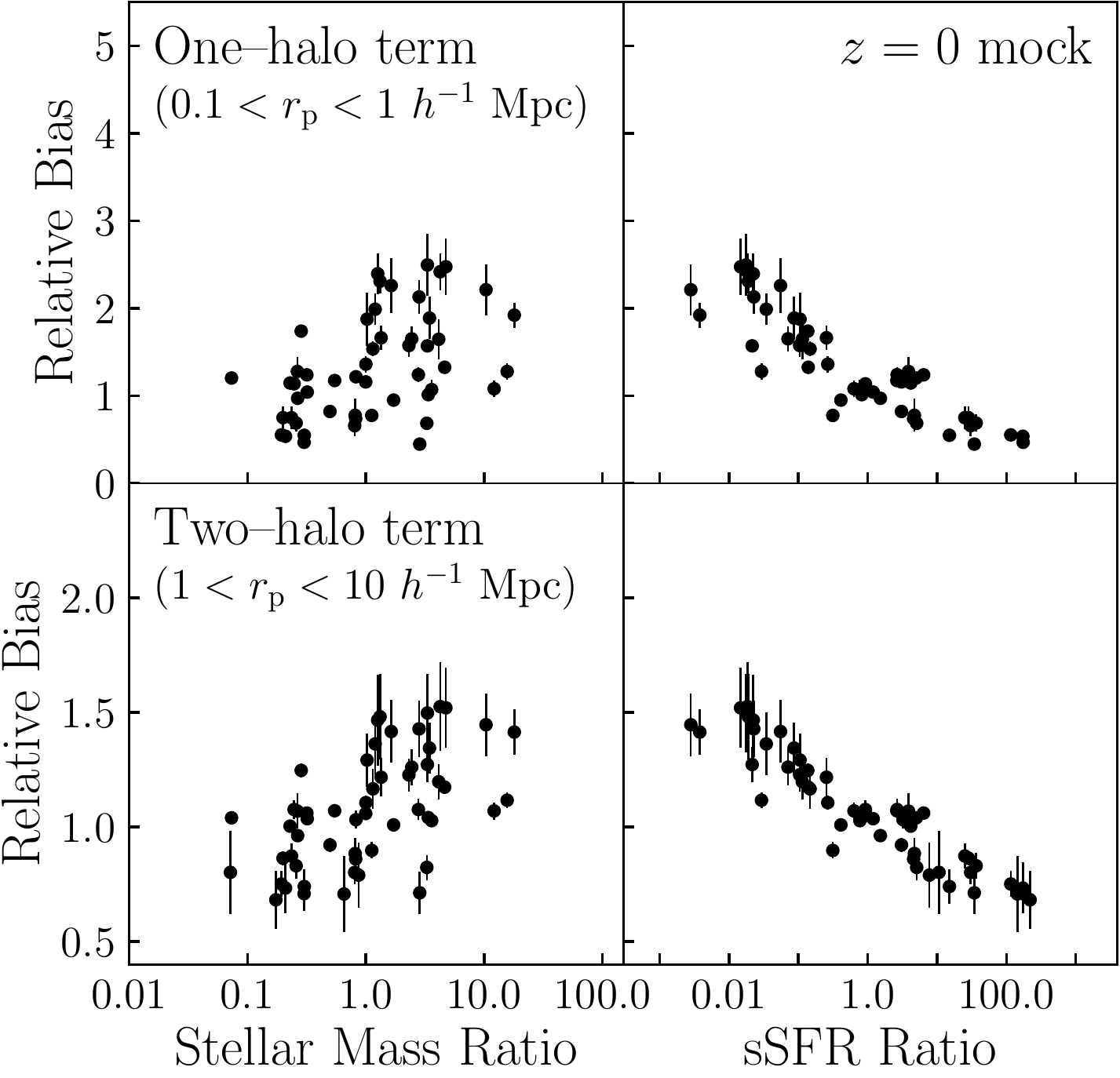} & &
\includegraphics[width=\linewidth]{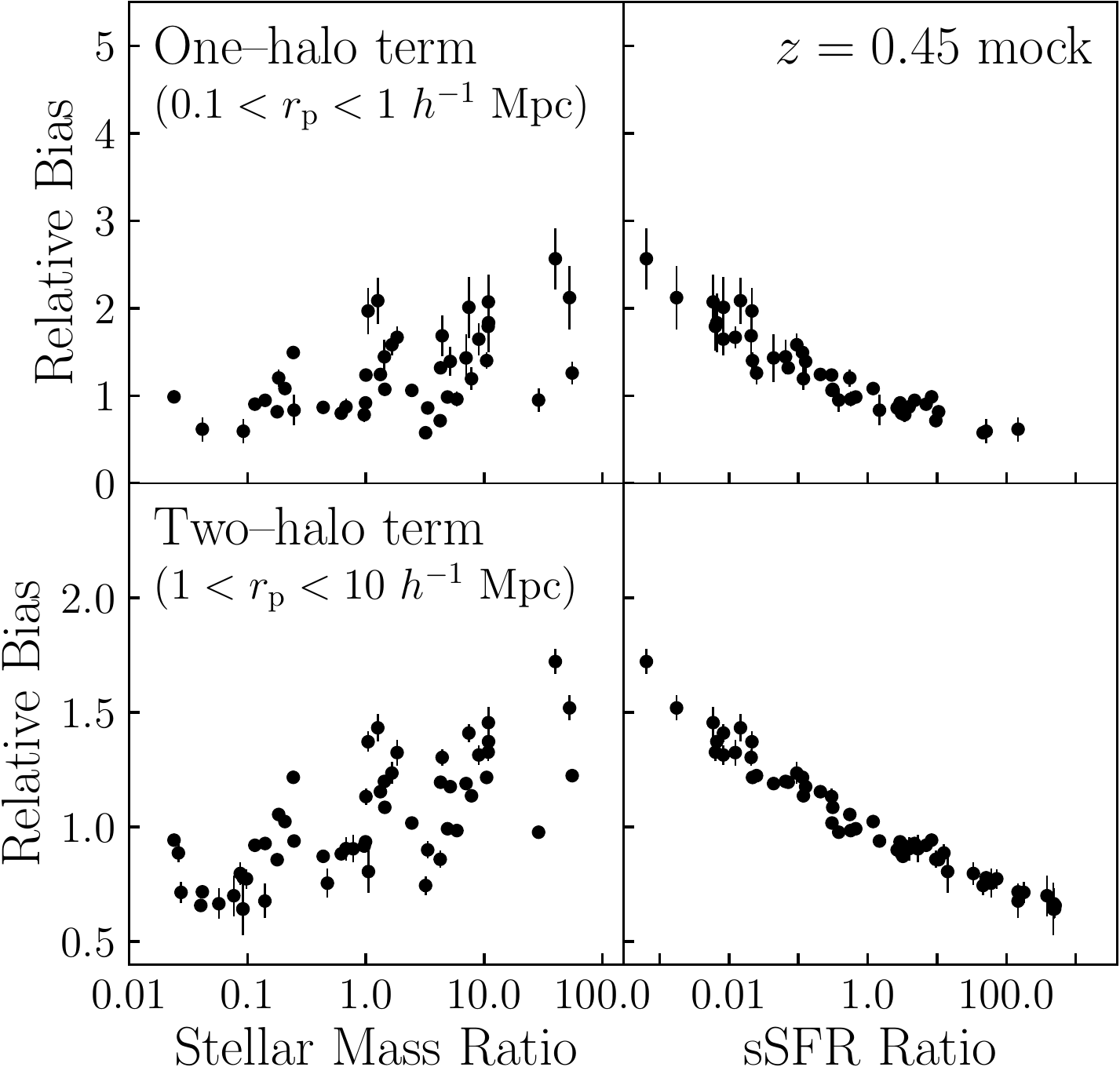} \\
\multicolumn{1}{c}{\textbf{(c)} $z=0$ mock} & &
\multicolumn{1}{c}{\textbf{(d)} $z=0.45$ mock} \\
\end{tabular}
\caption{
(a) Relative biases of pairs of SDSS galaxy samples as a function of each pair's stellar mass ratio (left column) and sSFR ratio (right column). The top and bottom rows show results one one-halo and two-halo scales, respectively.
(b) Relative biases of pairs of $z=0$ mock galaxy samples as a function of each pair's stellar mass ratio and sSFR ratio.
(c) Same as panel (a) but for PRIMUS galaxy samples at \zlow (recreated from data previously published in C17).
(d) Same as panel (b) but for the $z=0.45$ mock.
}
\label{fig04:rel_bias_2D}
\end{figure*}

We now compare the \emph{relative} biases between pairs of galaxy samples in SDSS and PRIMUS data to the corresponding galaxy sample pairs in mocks. The relative clustering strengths of galaxy samples within the same volume can have smaller uncertainty than the absolute biases because cosmic variance largely cancels out in relative bias measurements.

\citet{berti_etal19} refer to the clustering strength dependence on sSFR \emph{within} the star-forming or quiescent sequence as ``intra-sequence relative bias" (ISRB). We adopt that term here to refer to the relative biases of our ``main sequence split" galaxy samples in data and mocks.

The ISRB of quiescent galaxies in the $z=0$ mock agrees well with SDSS data within the errors:\ the relative bias between quiescent mock galaxies with above average versus below average sSFRs is ${1.24\pm0.03}$ on one-halo scales and ${1.13\pm0.03}$ on two-halo scales, which is $\sim9\%$ and $\sim4\%$ lower than in SDSS data, respectively.
This agreement disappears for the star-forming population, however, where in the $z=0$ mock star-forming galaxies with below average sSFRs are \emph{less} strongly clustered than those with above average sSFRs, the opposite of what we find in SDSS data. On both one-halo and two-halo scales the ISRB within the star-forming $z=0$ mock galaxy population is less than unity on both one-halo and two-halo scales, while it is greater than unity in the data.

At higher redshift there is good qualitative agreement between the mocks and PRIMUS data at both $z=0.45$ and $z=0.9$. The ISRB C17 find in quiescent and star-forming PRIMUS galaxies on both one-halo and two-halo scales, at both \zlow and \zhigh, is also present in the $z=0.45$ and $z=0.9$ mocks, although the \emph{magnitude} of the ISRB present in the mocks differs somewhat from the corresponding PRIMUS data.

In the $z=0.45$ mock the ISRB values are generally $\sim10\%$ lower than in the data, with the exception of the quiescent one-halo term, which is $\sim5\%$ greater in the mock than in the data. The ISRB in the $z=0.9$ mock is $\sim20\%$ to $40\%$ lower than in the data with the exception of the one-halo term for star-forming galaxies, which agrees precisely with the PRIMUS data value.

Following the presentation in C17, Figure~\ref{fig04:rel_bias_2D} shows the relative bias between pairs of galaxy samples in SDSS of PRIMUS data (panels (a) and (b),\footnote{Panel (b) of Figure~\ref{fig04:rel_bias_2D} is recreated with permission from results previously published in C17.} respectively), and between corresponding pairs of galaxy samples in the relevant mock (panels (c) and (d)), as functions of the stellar mass \emph{ratio} and sSFR \emph{ratio} of each pair of galaxy samples. In other words, the data points in Figure~\ref{fig04:rel_bias_2D} are a set of unique pairs taken from each of the galaxy samples we create in the stellar mass--sSFR plane. For example, the two ``star-forming/quiescent split" samples yield one pair:\ ``red/blue". Relative bias is shown versus stellar mass ratio on the left of each of the four panels, and verses sSFR ratio on the right. The top of each panel shows results on one-halo scales, while two-halo scales are shown on the bottom of each panel.

All possible galaxy sample pairs are not shown in panel (b) of Figure~\ref{fig04:rel_bias_2D} because C17 include only pairs of PRIMUS data samples for which the one-halo relative bias error is less than 25\%. In essence, the larger sample sizes provided by SDSS allow us to fill this parameter space more completely at $z\sim0$. We show results for $z\sim0$ and $z\sim0.45$; the results at $z\sim0.9$ are very similar.

We find similar trends at $z\sim0$ compared to C17's results at higher redshift. On both one-halo and two-halo scales the correlation between relative bias and sSFR ratio clearly has less scatter than the trend with stellar mass ratio.
A quick comparison of the the right columns of panels (a) and (b) may seem to suggest the correlation of relative bias with sSFR is tighter at higher redshift, but this could also be due to the lower signal-to-noise of PRIMUS data at \zlow compared to SDSS data. 

Panels (c) and (d) of Figure~\ref{fig04:rel_bias_2D} present the same analysis in the $z=0$ and $z=0.45$ mocks. The mocks at both redshifts display a similar relatively tight dependence of relative bias on sSFR ratio that we see in the data. There is little to no correlation between relative bias and stellar mass ratio at either redshift, in qualitative agreement with the $z\sim0$ data.
For clarity Figure~\ref{fig04:rel_bias_2D} does not show results for the $z=0.9$ data or mock, but the trends agree with our results at lower redshift.

\begin{figure*}[t]
\centering
\includegraphics[width=0.6\linewidth]{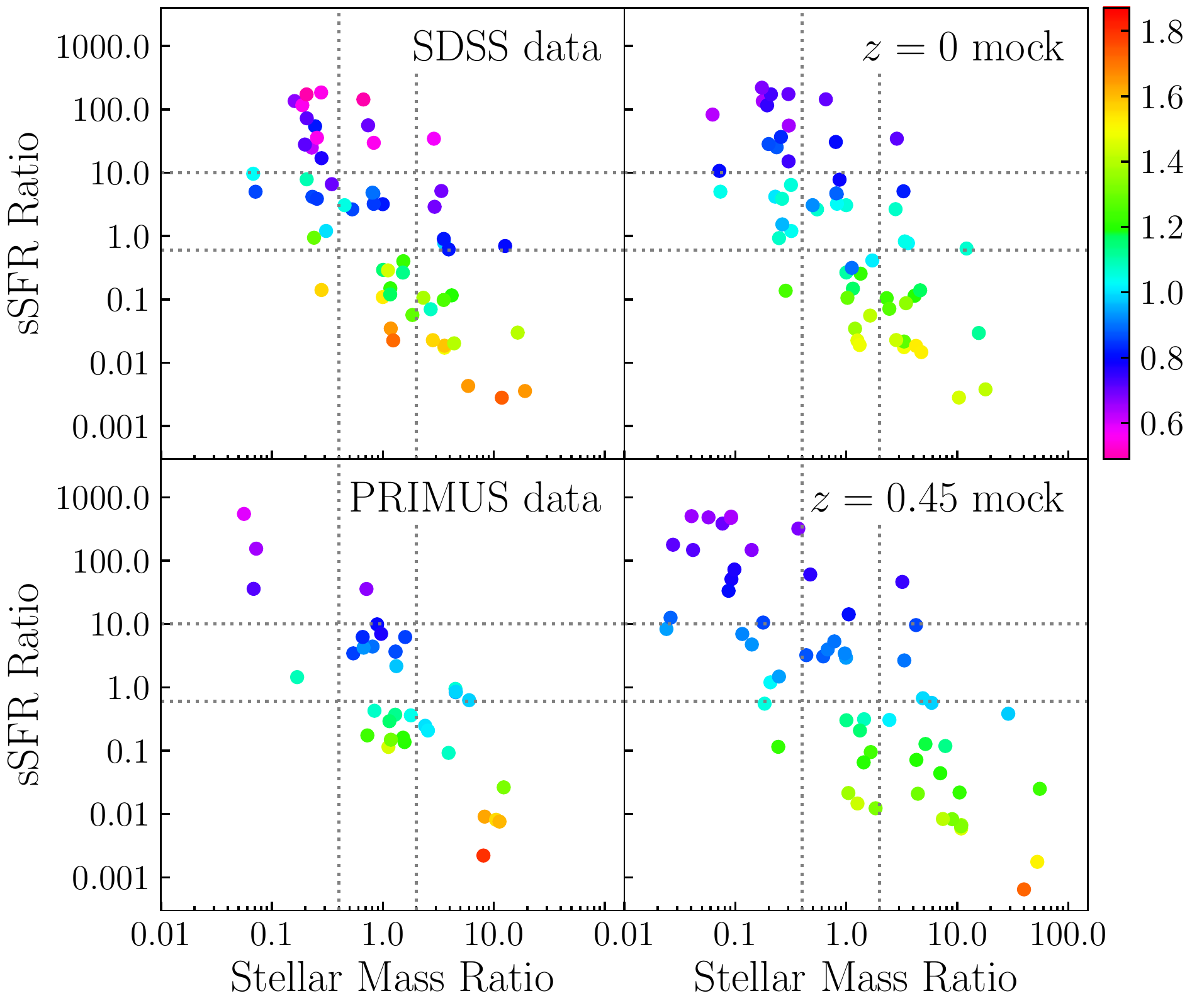}
\caption{
Two-halo relative bias of pairs of galaxy samples in the data (left column) and in the corresponding mock (right column) at $z=0$ (top row) and $z=0.45$ (bottom row), shown as a joint function of the stellar mass ratio and sSFR ratio of each pair of samples. The color of each point represents the magnitude of that pair's relative bias, as shown in the color bar. The dotted lines highlight regions of fixed stellar mass ratio or sSFR ratio where our galaxy samples probe several orders of magnitude in the ratio of the other parameter. Variations in sSFR ratio at fixed stellar mass ratio correspond to strong differences in relative bias, while variations in stellar mass ratio at fixed sSFR ratio do not result in substantially different relative bias values, demonstrating that galaxy clustering is a stronger function of sSFR than stellar mass.
}
\label{fig04:rel_bias_3D}
\end{figure*}

\subsection{Joint Dependence of Relative Bias on Stellar Mass and sSFR}\label{subsec04:rel_bias_joint}

Again following C17, in Figure~\ref{fig04:rel_bias_3D} we present a way to understand the \emph{joint} dependence of relative bias on both stellar mass ratio and sSFR ratio at in both the data and corresponding mocks. 
Figure~\ref{fig04:rel_bias_3D} shows the relative bias on two-halo scales of pairs of galaxy samples in the data (left column) and corresponding mock (right column) as a joint function of each pair's stellar mass ratio and sSFR ratio, with the magnitude of the relative bias represented by the color bar in the figure. The dotted lines bracket regions of either fixed stellar mass ratio or sSFR ratio where our samples probe several of magnitude in the ratio of the other parameter. The vertical dotted lines highlight sample pairs with stellar mass ratios of 0.6--2 and sSFR ratios from $\sim10^{-3}$ to $\sim10^2$. The relative biases of these sample pairs span the full range of relative bias values observed, from $\sim1.6$ for the smallest sSFR ratios to $\sim0.6$ for the largest sSFR ratios.
For comparison, the horizontal dotted lines highlight sample pairs with sSFR ratios of 0.9--10, across three orders of magnitude in stellar mass ratio, and show little variation in relative bias across the range of stellar mass ratios probed by our samples.
These results confirm at $z\sim0$ the trends observed by C17 in PRIMUS data at ${0.2 < z < 1.2}$, namely that sSFR ratio is more relevant for determining the relative bias of two galaxy samples than stellar mass ratio. This implies that clustering amplitude is more fundamentally linked to sSFR than to stellar mass.

\section{Modifying the Mock Galaxy Catalog at Low Redshift}\label{sec04:z0_mod}

As noted in \S\ref{subsec04:abs_bias} there is less agreement between the data and corresponding mock at $z\sim0$ than at either $z\sim0.45$ or $z\sim0.9$, particularly within the SFMS, where sSFR in the $z=0$ mock is anticorrelated with clustering strength (left panels of Figure~\ref{fig04:abs_bias_data_vs_mocks}). This anticorrelation does \emph{not} appear in the data at any of the redshifts studied here, nor is it present in the $z=0.45$ or $z=0.9$ mock. 

This arises in the \UM because there is a redshift-dependent parameter ($\sigma_\mathrm{SF,uncorr}(z)$) that determines the scatter between SFR in the main sequence and halo accretion rate. 
Specifically, the proxy used for halo accretion rate is $\dvmax,$ the change in the maximum circular velocity of a halo over the last dynamical time, ${\tau_{\rm dyn}\equiv \left(\sfrac{4}{3}\,\pi\,G\,\rho_{\rm vir}\right)^{\sfrac{\!\!-\!\!1}{\,2}}}$, where $\rho_{\rm vir}$ is the virial overdensity \citep{bryan_norman98}.\footnote{In detail, we use the larger of $\tau_{\rm dyn}$ or $\tau_{\rm M_{peak}},$ the time at which the halo reached its peak mass, to account for sustained quenching of extremely stripped (sub)halos. See \S3.1 of \citet{behroozi_etal19} for additional details.}  

The $\sigma_\mathrm{SF,uncorr}(z)$ parameter has no direct constraints from any of the observations used as inputs to the \UM, however, and so functions as a nuisance parameter to capture uncertainty in the underlying galaxy--halo relationship.  In the current best-fit \UM model, this parameter apparently results in a moderate correlation strength between main sequence SFR and halo accretion at $z\gtrsim 0.5$ and a negligible correlation strength at $z=0$, causing the behavior seen in Figure~\ref{fig04:abs_bias_data_vs_mocks}.

Evidently, the new SDSS measurements presented here have the constraining power to directly measure this previously-unknown correlation strength. Motivated by this additional constraining power, we examine how the strength of this galaxy-halo correlation can be better understood through the full, two-dimensional dependence of clustering upon stellar mass and sSFR.

Rather than rerunning the full \UM machinery while including these new measurements, which is beyond the scope of this paper, here we carry out a targeted study of this effect by creating a ``modified" $z=0$ mock with an imposed stronger correlation between galaxy SFR and halo mass accretion rate (using \dvmax at fixed \vmax at $M_{\rm peak}$\footnote{\vmax at $M_{\rm peak}$ is the maximum circular halo velocity at $M_{\rm peak}$, the peak historical mass achieved by a halo.} as a proxy). We created this mock by re-running the \UM with the best-fit parameter set, except that we lowered $\sigma_\mathrm{SF,uncorr}$ at $z=0$ so that the correlation between main sequence SFR and halo accretion was effectively the same ($r=0.5$) as at higher redshifts.

\begin{figure*}[t]
\centering
\includegraphics[width=0.9\linewidth]{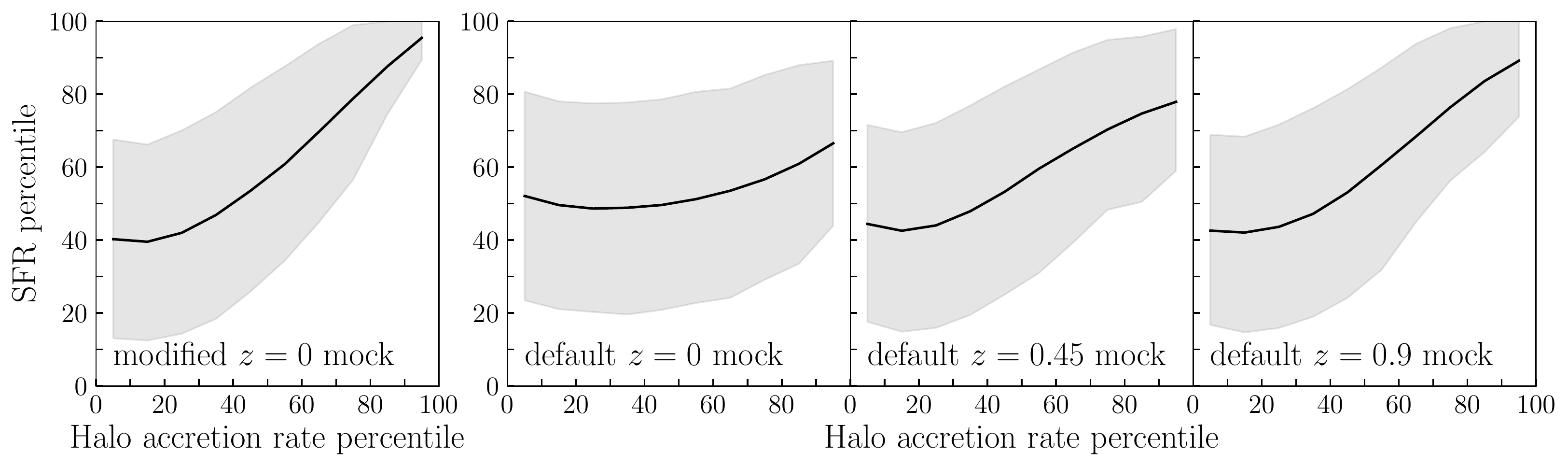}
\caption{
Mean galaxy SFR percentile versus halo accretion rate percentile for star-forming mock galaxies in the modified $z=0$ mock (left panel) and the default mocks at $z=0$, $0.45$, and $0.9$ (right three panels). In the ``modified" $z=0$ mock, the correlation between SFR and halo accretion rate for star-forming galaxies is increased to $\sim0.5$ from the default value of $\sim0$. The shaded gray regions show $1\sigma$ deviation on the mean SFR.
}
\label{fig04:sfr_dvmax}
\end{figure*}

Figure~\ref{fig04:sfr_dvmax} demonstrates this modification visually, showing the SFR versus halo accretion rate\footnote{Specifically, Figure~\ref{fig04:sfr_dvmax} shows the relationship between rank-ordered SFR and rank-order halo accretion rate.} for star-forming galaxies in the default mocks at $z=0$, 0.45, and 0.9, and for the ``modified" $z=0$ mock. In the default model the SFR--halo accretion rate correlation for star-forming galaxies is strongest in the $z=0.9$ mock and declines to a very shallow trend at $z=0$. In our ``modified" $z=0$ mock this correlation is boosted to be as strong as it is at higher redshift in the default model.

Boosting the SFR--halo accretion rate correlation also affects the distribution of star-forming satellite galaxies within the SFMS. As listed in Table~\ref{tab04:samples_mock} above, the ``main sequence split" galaxy samples in the default $z=0$ mock have satellite fractions of 0.26 for star-forming galaxies above the SFMS, and 0.24 for star-forming galaxies below the SFMS. In the modified $z=0$ mock these satellite fractions are 0.20 and 0.28 for star-forming galaxies above and below the SFMS, respectively.

\begin{figure*}[t]
\centering
\setlength{\tabcolsep}{0pt}
\begin{tabular}{ p{0.4\textwidth} p{0.05\textwidth} p{0.55\textwidth} }
\includegraphics[width=\linewidth]{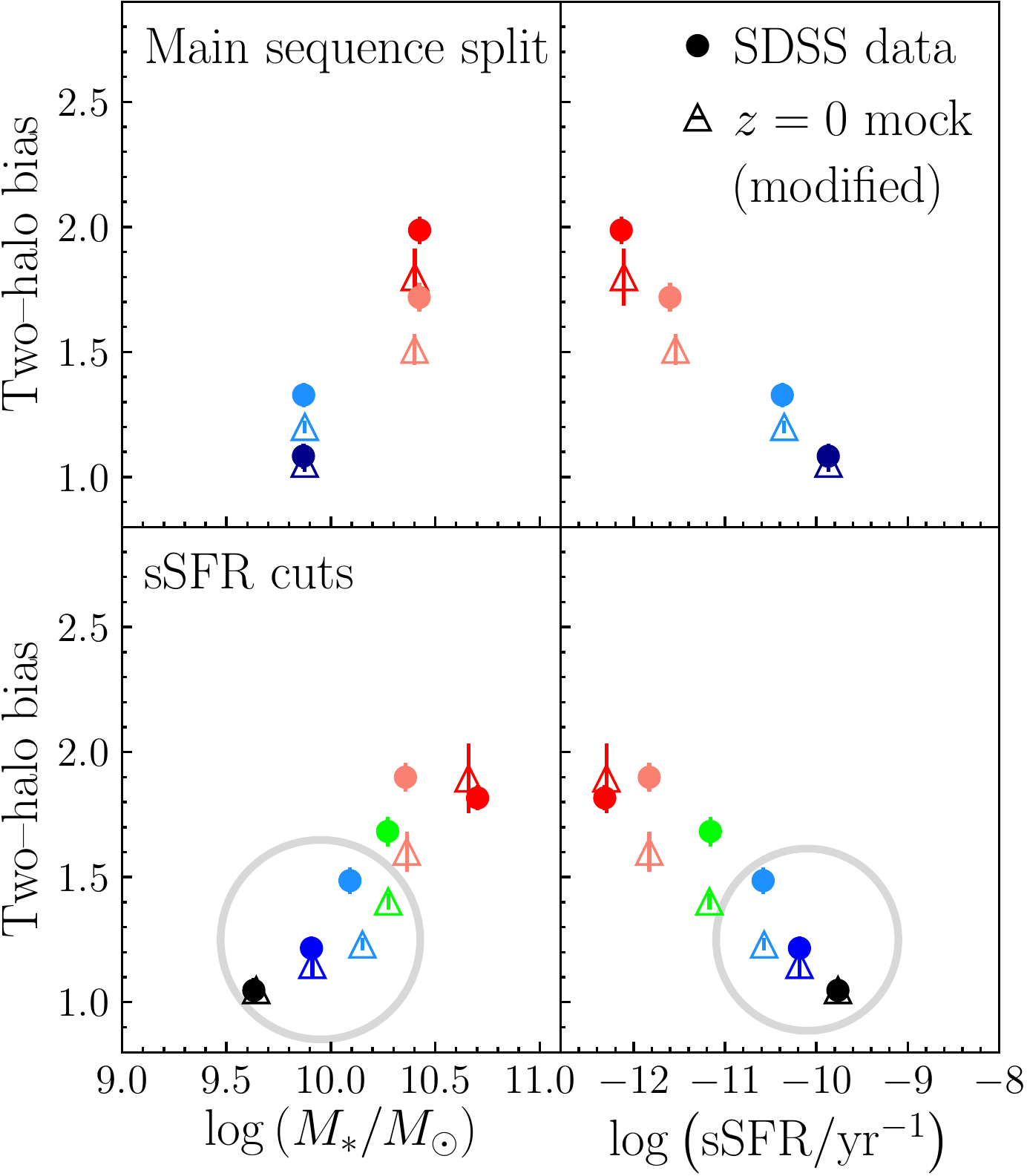} & &
\includegraphics[width=\linewidth]{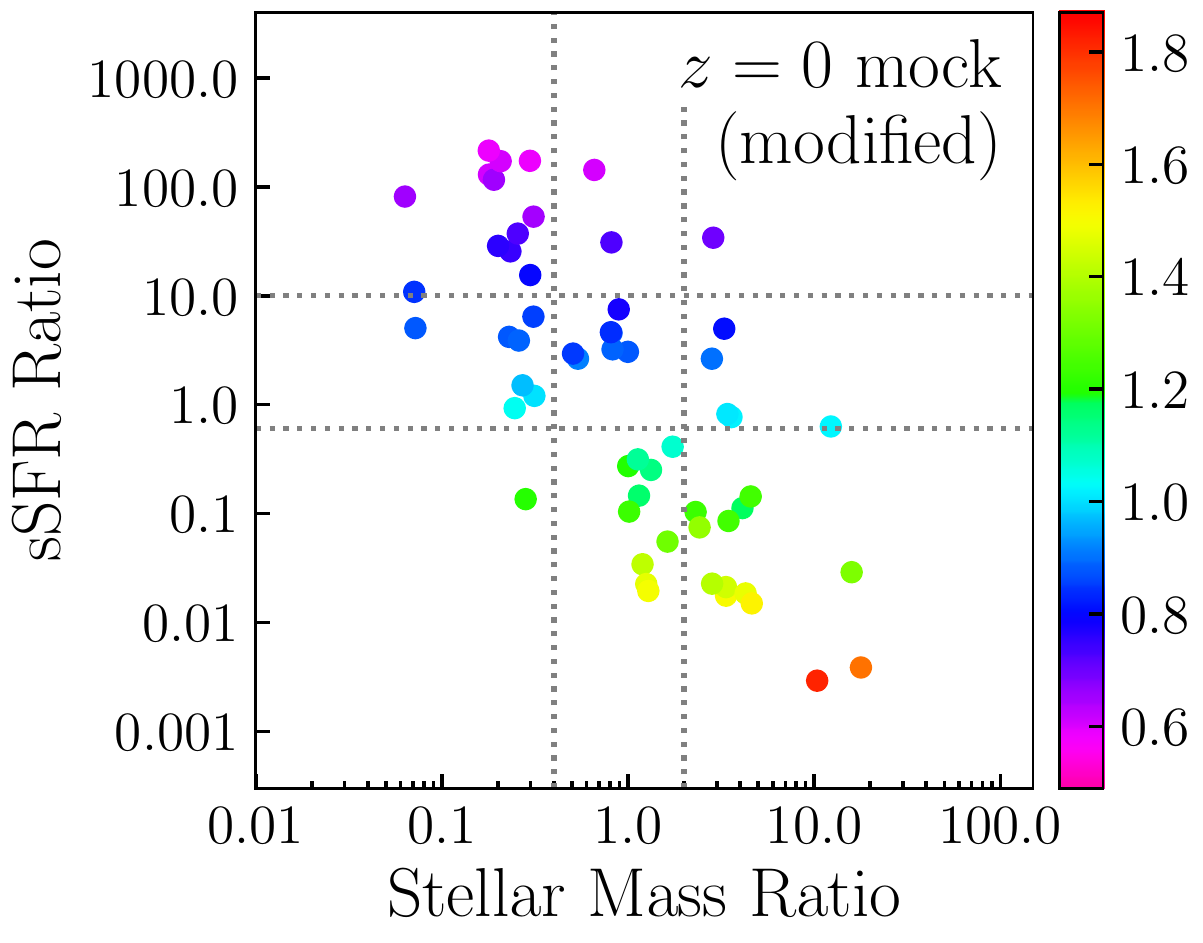} \\
\end{tabular}
\caption{
Left:\ Same as the left four panels of Figure~\ref{fig04:abs_bias_data_vs_mocks} but for the ``modified" $z=0$ mock.
The gray circles both here and in Figure~\ref{fig04:abs_bias_data_vs_mocks} above highlight the difference between the ``default'' and ``modifed'' $z=0$ mocks, and clearly show that the ``modified'' $z=0$ mock reproduces galaxy bias as a function of sSFR for star-forming SDSS galaxies better than the default model.
Right:\ Same as the upper right panel of Figure~\ref{fig04:rel_bias_3D}, but for the ``modified" $z=0$ mock. Shown is the relative bias on two-halo scales of mock galaxy sample pairs as a joint function of stellar mass ratio and sSFR ratio; the color of each point indicates the relative bias value.
The correlation between relative bias and sSFR ratio at fixed stellar mass is stronger in the ``modified" mock relative to the default $z=0$ mock; this modification also improves the fit to SDSS data (upper left panel of Figure~\ref{fig04:rel_bias_3D}).
}
\label{fig04:z0p0_corr0p5}
\end{figure*}

\begin{figure*}[t]
\centering
\includegraphics[width=\linewidth]{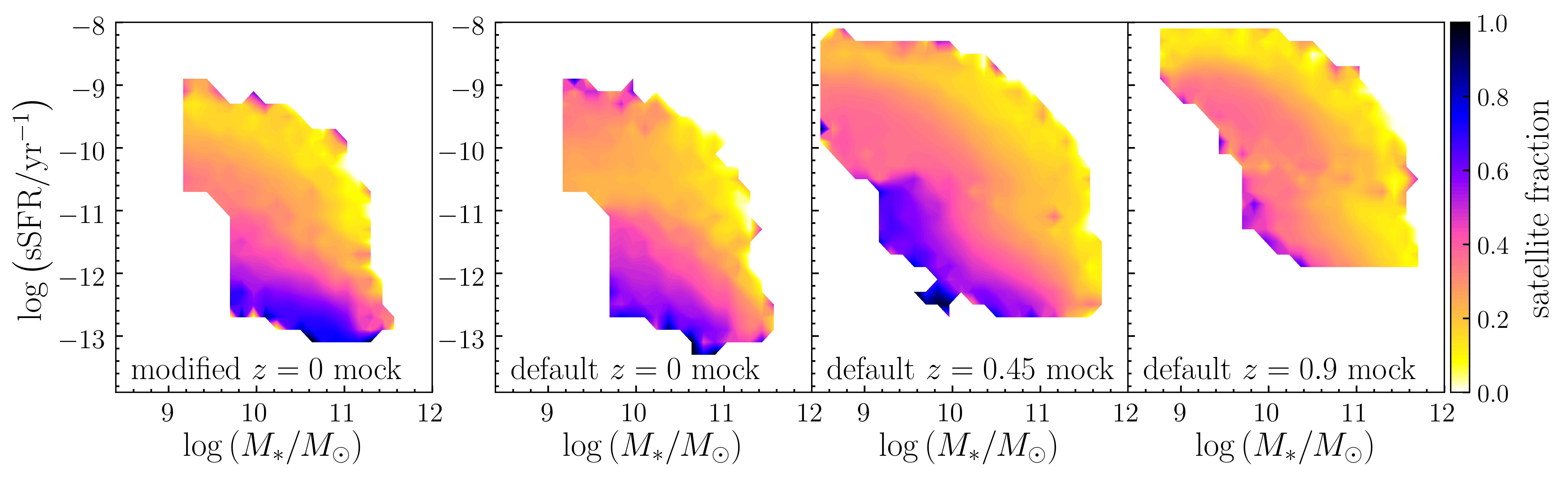}
\caption{
Satellite galaxy fraction as a function of stellar mass and sSFR for the ``modified" $z=0$ mock (left panel) and the default mocks at $z=0$, 0.45, and 0.9 (right three panels). The modified $z=0$ mock brings the satellite fraction distribution into closer agreement with the default mocks at higher redshift.
}
\label{fig04:sat_fracs}
\end{figure*}

In Figure~\ref{fig04:z0p0_corr0p5} we repeat the analysis of Figures~\ref{fig04:abs_bias_data_vs_mocks} and \ref{fig04:rel_bias_3D} for the ``modified" $z=0$ mock. The left panels show absolute bias on two-halo scales as a function of mean stellar mass and mean sSFR for SDSS data and the ``modified" $z=0$ mock for the ``main sequence split" (top row) and ``sSFR cuts" (bottom row) sets of galaxy samples split into three bins in sSFR. As in Figure~\ref{fig04:abs_bias_data_vs_mocks}, gray circles highlight bias values for star-forming samples. Comparing the gray circled regions in Figure~\ref{fig04:abs_bias_data_vs_mocks} to those in Figure~\ref{fig04:z0p0_corr0p5} clearly shows the ``modified" $z=0$ mock agrees with SDSS data much better than the default model. Although the slope is shallower than in the data, in the ``modified'' $z=0$ mock the bias of star-forming galaxies increases with decreasing sSFR. (The trend for quiescent galaxies in unchanged between the ``default'' and ``modified" $z=0$ mocks.)

The right panel of Figure~\ref{fig04:z0p0_corr0p5} is the same as the upper right panel of the Figure~\ref{fig04:rel_bias_3D}, but for the ``modified" $z=0$ mock:\ two-halo relative bias of galaxy sample pairs as a joint function of the stellar mass ratio and sSFR ratio of each pair. The difference between the default and ``modified" $z=0$ mocks is the range of relative bias values. In the default $z=0$ mock the two-halo relative bias varies from 0.63 at the smallest sSFR ratios to 1.53 at the largest. This expands to 0.58--1.81 in the ``modified" $z=0$ mock over the same range of sSFR ratios, which is a better match to the range seen in SDSS data of 0.49--1.73. These results highlight the constraining power of both the absolute bias as a function of sSFR and the relative bias as a function of sSFR ratio for pairs of galaxy samples in constraining empirical models of galaxy evolution. In the following section we use exclusively the ``modified" $z=0$ mock and the default mocks at $z=0.45$ and $z=0.9$.

\section{Contribution of Central Galaxies to Intra-sequence Relative Bias}\label{sec04:cen_sat}

In this section we use the mock galaxy catalogs to investigate the relative contributions of central and satellite galaxies to the result of \S\ref{sec04:wp_bias} that galaxy clustering is a stronger function of sSFR at a given stellar mass than of stellar mass at fixed sSFR.
Here we use exclusively the ``modified" $z=0$ mock and the default mocks at $z=0.45$ and $z=0.9$.

\subsection{Relative Bias of Mock Centrals and Satellites}\label{subsec04:rel_bias_cen_sat}

\begin{figure*}[t]
\normalsize
\centering
\setlength{\tabcolsep}{0pt}
\begin{tabular}{p{0.4275\textwidth} p{0.0225\textwidth} p{0.4275\textwidth}}
\includegraphics[width=\linewidth]{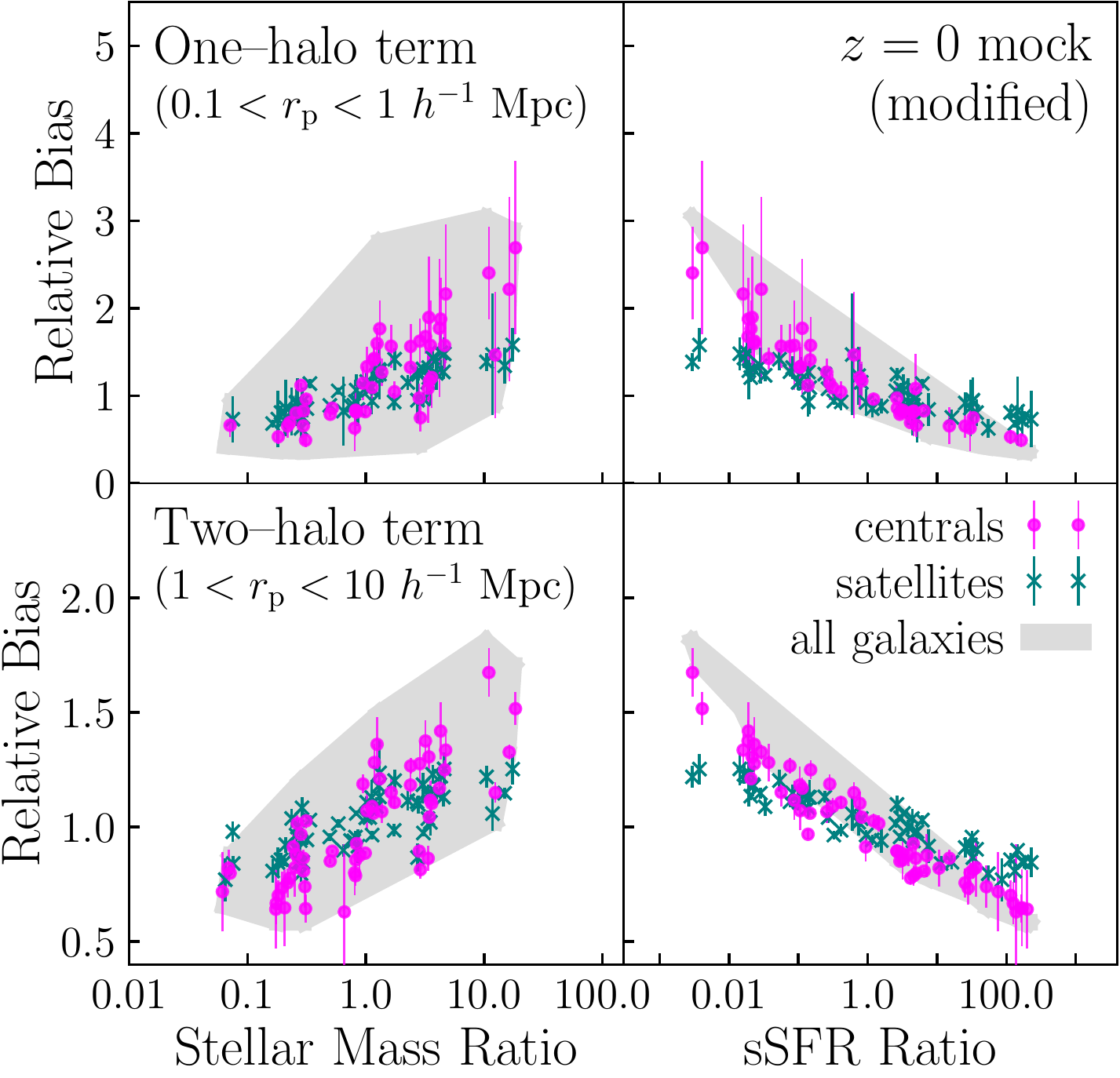} & & 
\includegraphics[width=\linewidth]{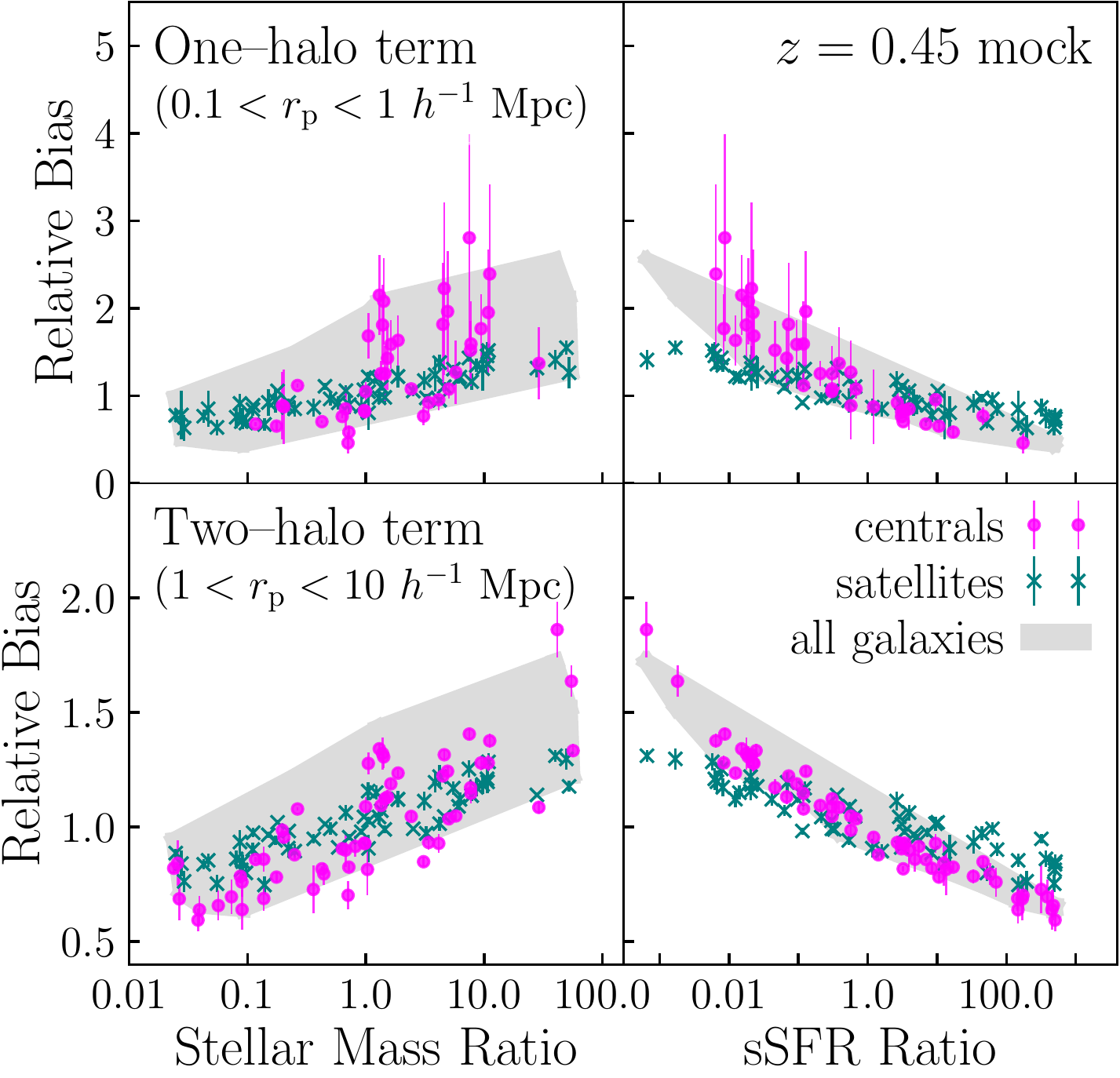} \\
\multicolumn{1}{c}{\textbf{(a)} $z=0$ mock} & &
\multicolumn{1}{c}{\textbf{(b)} $z=0.45$ mock} \\
\end{tabular}
\caption{
(a) Relative bias between pairs of mock galaxy samples at $z=0$ as a function of each pair's stellar mass ratio and sSFR ratio, divided into centrals (magenta circles) and satellites (green $\times$ symbols). The grey shaded region is the result for all mock galaxies (i.e.\ without distinguishing centrals from satellites, as shown in panels (c) and (d) of Figure~\ref{fig04:rel_bias_2D}.) For clarity only points with errors less than 50\% of the relative bias value are shown.
(b) Same as panel (a) but for the $z=0.45$ mock. Results at $z=0.9$ are not shown but exhibit the same trends as those seen at lower redshifts:\ the correlations between relative bias and both sSFR and stellar mass ratio are due predominantly to central galaxies.
}
\label{fig04:rel_bias_2D_cen_sat}
\end{figure*}

\begin{figure*}[t]
\centering
\includegraphics[width=0.6\linewidth]{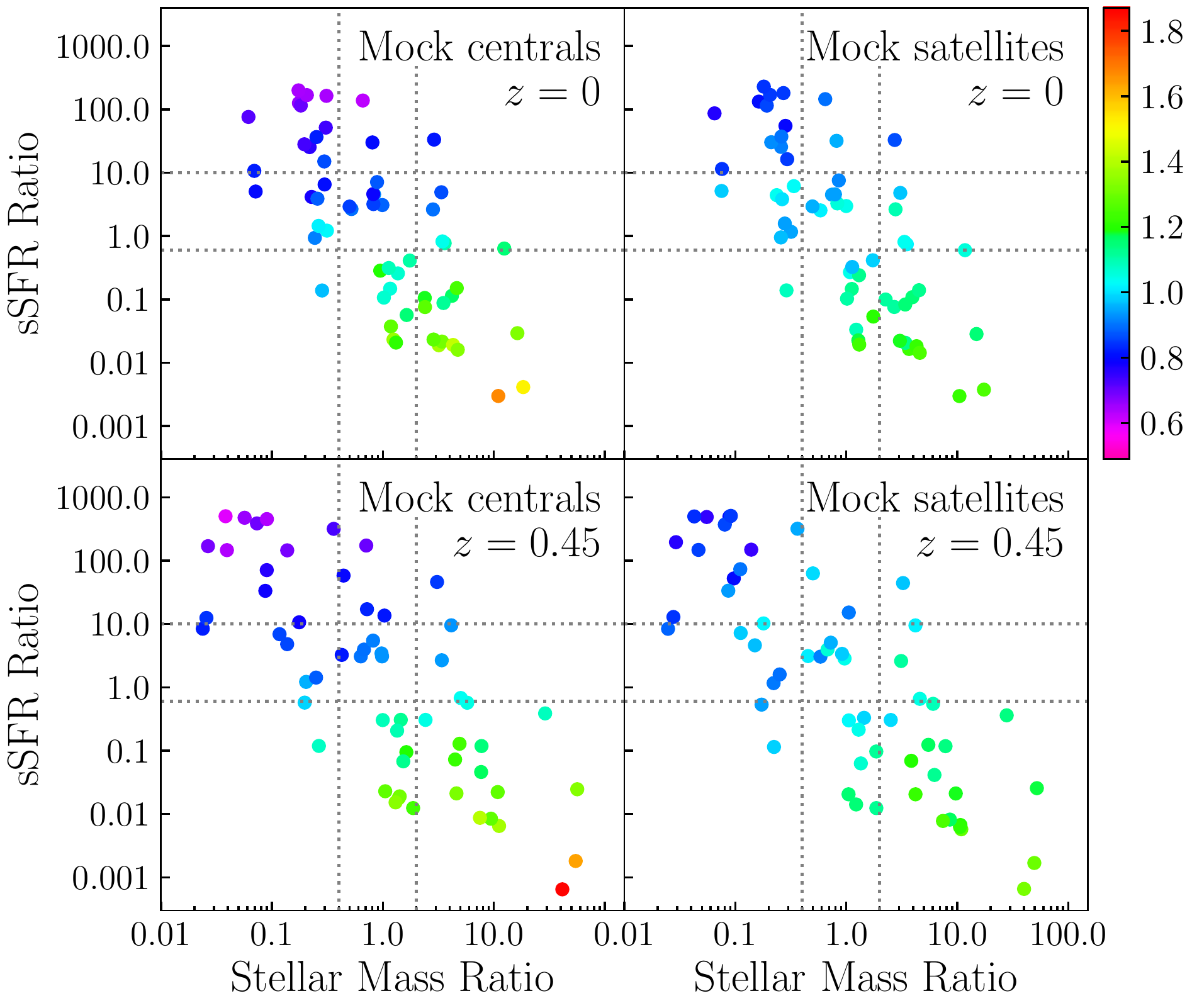}
\caption{
Similar to Figure~\ref{fig04:rel_bias_3D} for the $z=0$ (top row) and $z=0.45$ mocks (bottom row):\ two-halo relative bias of mock galaxy sample pairs, shown as a joint function of stellar mass ratio and sSFR ratio, and divided into central and satellite galaxies (left and right columns, respectively).
For clarity results for the $z=0.9$ mock are not shown, but follow the same trends seen in the $z=0$ and $z=0.45$ mocks.
Dotted lines highlight regions of fixed stellar mass ratio or sSFR ratio where our galaxy samples probe several orders of magnitude in the ratio of the other parameter. 
The dependence of relative bias on sSFR ratio at fixed stellar mass ratio (and not vice versa) is due predominantly to centrals; pairs of central galaxy samples span a larger range of relative biases compared to the range for satellite galaxies over the same span of sSFR ratios.
}
\label{fig04:rel_bias_3D_cen_sat}
\end{figure*}

Having shown agreement between data and mocks when considering various galaxy samples drawn from the full galaxy population, we now divide each mock galaxy sample into central and satellite components to determine the contribution of each to the trends seen in Figures~\ref{fig04:rel_bias_2D} and \ref{fig04:rel_bias_3D}.
Figure~\ref{fig04:sat_fracs} shows how the satellite galaxy fraction in the mocks changes with both stellar mass and sSFR. Like the presentation in Figure~\ref{fig04:sfr_dvmax}, the set of three panels on the right shows the default mocks, while the leftmost panel shows the ``modified" $z=0$ mock. The satellite fraction distribution is qualitatively different for star-forming galaxies in the default $z=0$ mock compared to the default at higher redshift. In the ``modified" $z=0$ the satellite fraction distribution more closely resembles that of the higher redshift default mocks:\ within the SFMS the satellite fraction is lowest ($\lesssim10\%$) for the highest sSFR galaxies and increases fairly smoothly across the main sequence to $\sim40\%$ for the lowest sSFR star-forming galaxies. In the default $z=0$ mock this trend is reversed across the lower-mass end of the SFMS.

Results showing the dependence of the relative bias as a function of stellar mass ratio and sSFR ratio separated in the mock samples into centrals and satellites are shown in Figure~\ref{fig04:rel_bias_2D_cen_sat}; centrals are represented by magenta circles and satellites by green $\times$ symbols. The gray shaded region is the result for all mock galaxies shown in panels (c) and (d) of Figure~\ref{fig04:rel_bias_2D}.
We focus on the two-halo relative bias (bottom row of Figure~\ref{fig04:rel_bias_2D_cen_sat}) as this is the length scale at which central galaxy relative bias is well defined, and therefore also the scale at which meaningful comparisons can be made between centrals and satellites. The same trends are seen on smaller scales (top row), but the errors are larger for the central galaxy relative bias. 

In both the $z=0$ and $z=0.45$ mocks the correlation between relative bias and sSFR ratio is driven by central galaxies. This can be clearly seen in the lower right sections of both panels (a) and (b) of Figure~\ref{fig04:rel_bias_2D_cen_sat}.
In the $z=0$ mock the two-halo relative bias of central galaxies increases from $\sim0.6$ to $\sim1.7$ with decreasing sSFR ratio across the nearly five orders of magnitude in sSFR ratio probed here. 
Over the same range in sSFR ratio the two-halo relative bias of mock satellites spans the narrower range of $\sim0.8$ to $\sim1.3$.
At $z=0.45$ the two-halo relative bias of mock centrals is again anticorrelated with sSFR ratio across five orders of magnitude, from $\sim0.6$ at the largest sSFR ratios to $\sim1.8$ at the smallest sSFR ratios (lower right of panel (b) in Figure~\ref{fig04:rel_bias_2D_cen_sat}). There is a shallower anticorrelation with slightly larger scatter between relative bias and sSFR ratio for mock satellites over the same range of sSFR ratios.
We also note that in stellar mass ratio space the division between central and satellite galaxies is not as clean as in sSFR ratio space.

Figure~\ref{fig04:rel_bias_3D_cen_sat} illustrates the difference between the trends for mock centrals and satellites differently (similar to the presentation of Figure~\ref{fig04:rel_bias_3D}), showing the relative bias of pairs of central galaxy samples (left column) and pairs of satellite galaxy samples (right column) as a joint function of each pair's stellar mass ratio and sSFR ratio.
In other words, to make this figure we remeasure the clustering and bias of each mock galaxy sample, first keeping only the central galaxies in each sample, then keeping only the satellites. We then calculate the relative bias of pairs of central-only samples, and likewise for satellite-only samples.\footnote{The stellar mass and sSFR ratios of both sets of samples are the same by design, as centrals and satellites are selected from the same parent samples (see Table~\ref{tab04:samples_mock}).}

The top and bottom rows of Figure~\ref{fig04:rel_bias_3D_cen_sat} show results for $z=0$ and $z=0.45$ mocks, respectively. At both $z=0$ and $z=0.45$ the relative biases of central galaxies span nearly the entire range of the color bar, from magenta at the low end ($\sim0.6$) to orange near the high end ($\sim1.8$). In contrast, the relative biases of satellites at both redshifts span a narrower section of the color bar, from dark blue ($\sim0.8$) to light green ($\sim1.3$).
We do not include the $z=0.9$ mock in Figure~\ref{fig04:rel_bias_3D_cen_sat}, but the results follow the same trends seen in mocks at lower redshift.

In summary, we find that sSFR ratio---not stellar mass ratio---is the key factor influencing the clustering amplitude differences between galaxy samples. Moreover, analysis of the \UM model indicates that this relationship is driven primarily by central galaxies.

\subsection{Intra-sequence Clustering Differences}\label{subsec04:isrb_cen_sat}

\begin{figure*}[h]
\centering
\setlength{\tabcolsep}{0pt}
\begin{tabular}{p{0.2925\textwidth} p{0.0025\textwidth} p{0.6775\textwidth}}
\includegraphics[width=\linewidth]{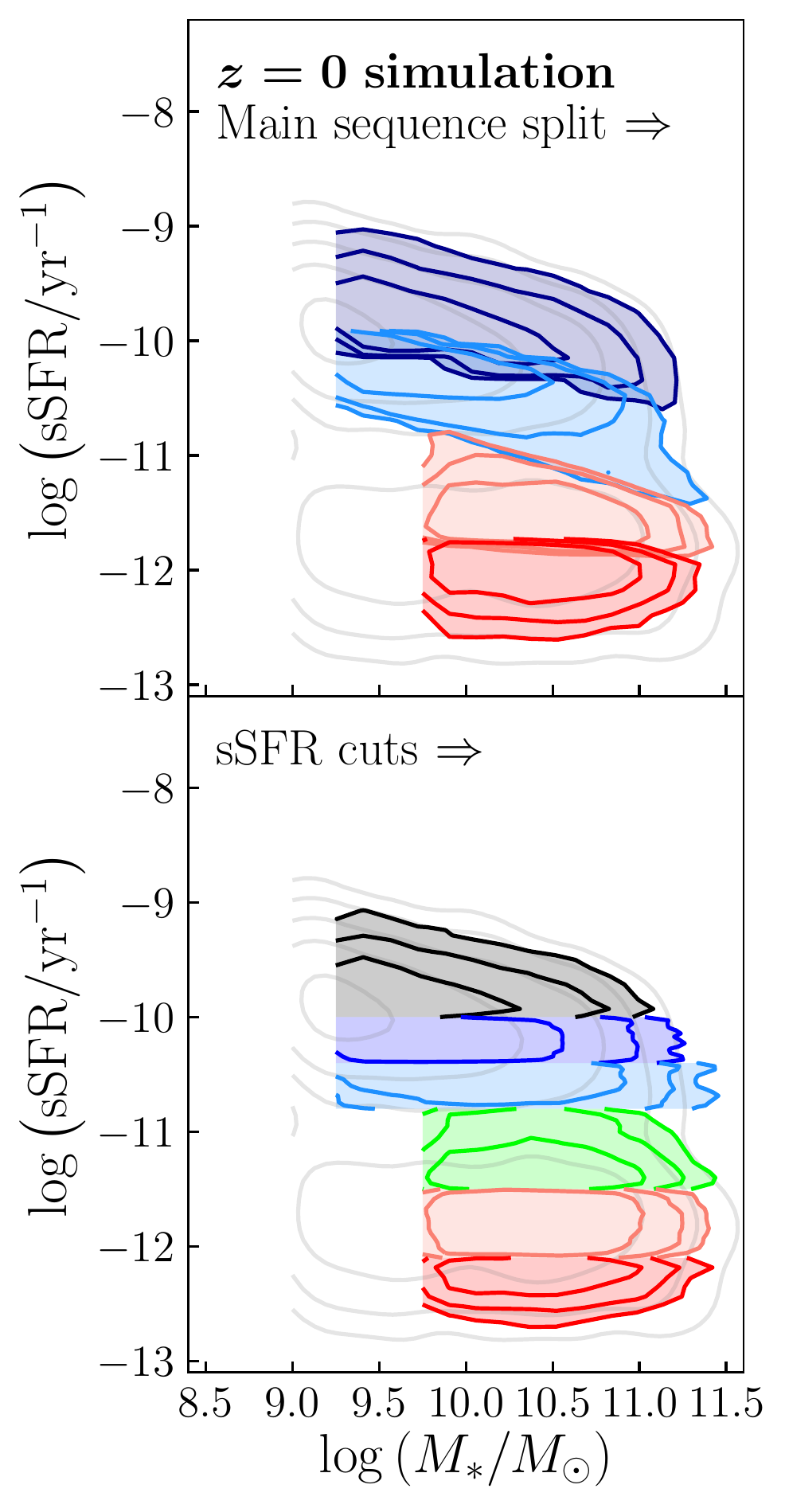} & &
\includegraphics[width=\linewidth]{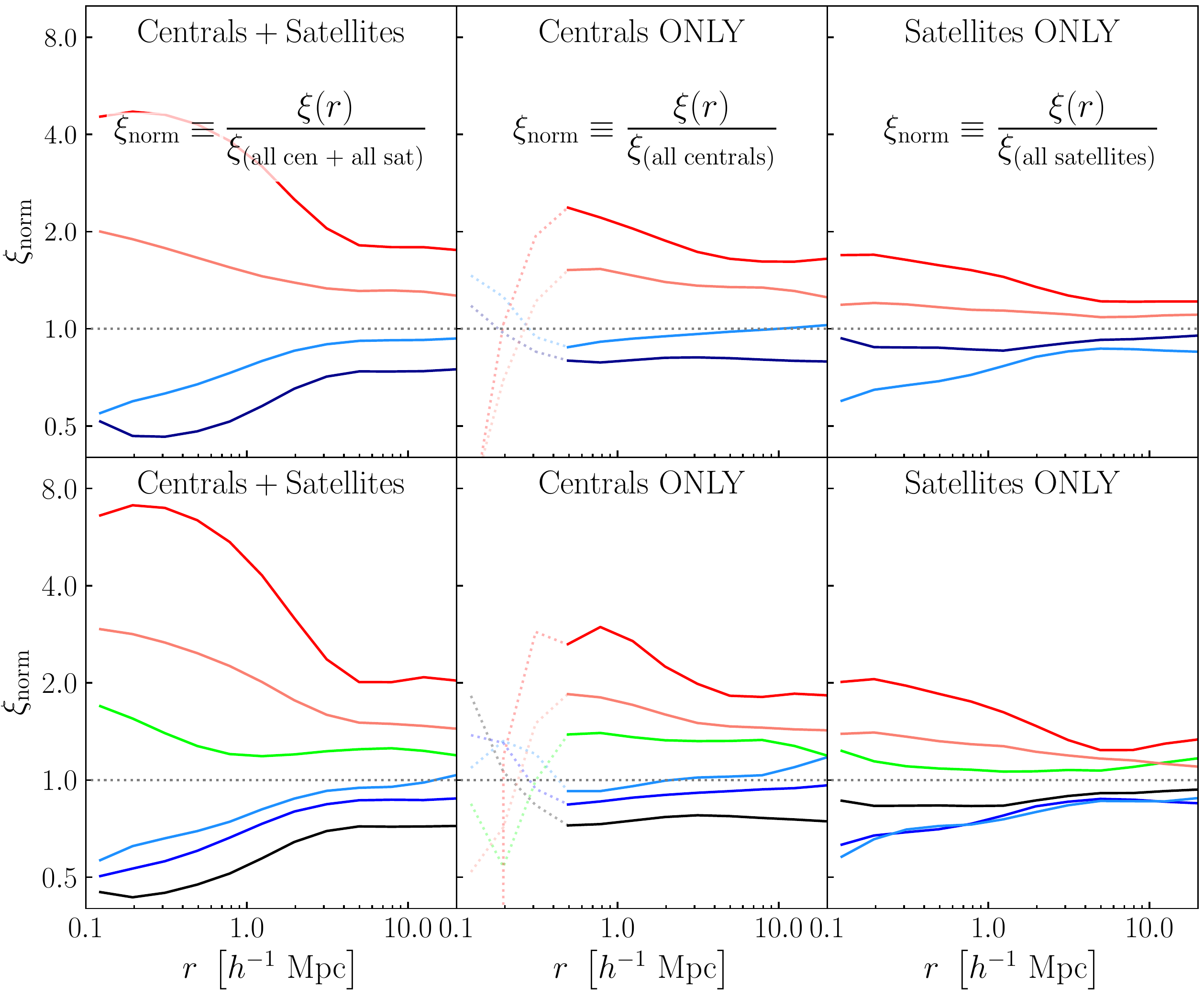} \\
\includegraphics[width=\linewidth]{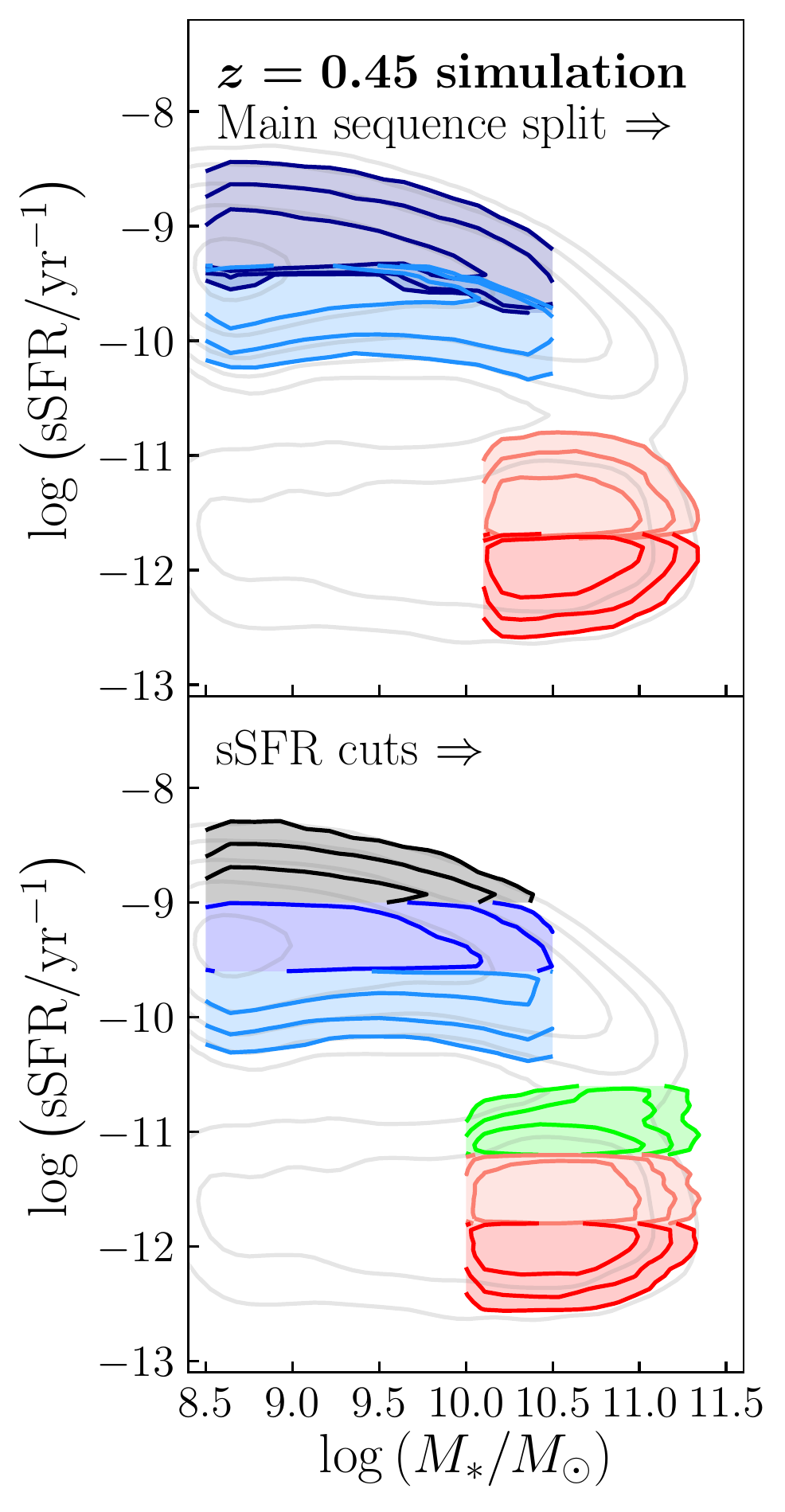} & &
\includegraphics[width=\linewidth]{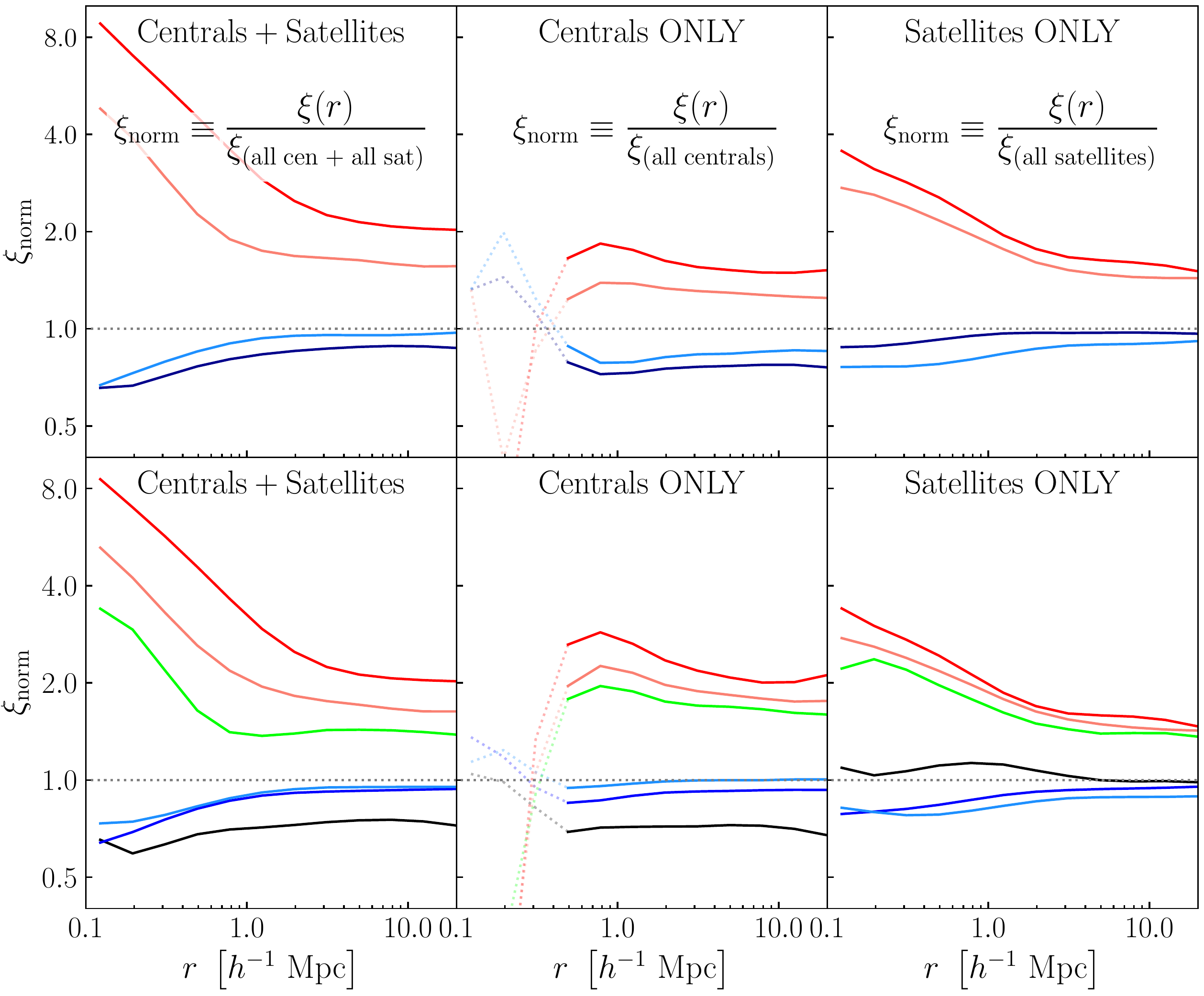} \\
\end{tabular}
\caption{Normalized \xir for the ``main sequence split" and ``sSFR cuts" samples selected from the $z=0$ (top two rows) and $z=0.45$ (bottom two rows) simulations. For each sample $\xi_{\rm norm}$ is \xir for that sample divided by the correlation function for all simulated galaxies \emph{of the relevant galaxy type}, i.e.\ centrals $+$ satellites (second column), centrals only (third column), or satellites only (last column), as indicated in the relevant panels. The first column shows the stellar mass--sSFR distributions of the relevant samples in each row.
While the stellar mass and sSFR cuts defining these samples are identical to those used to select mock galaxy samples used in previous sections, the samples shown here are taken from the full \UM model output, i.e.\ these samples are \emph{not} subject to observational effects such as magnitude limits or redshift errors.
}
\label{fig04:xi_norm}
\end{figure*}

Here we use the simulations from which the mock catalogs are drawn to investigate the extent to which the ISRB is due to central galaxies alone, as opposed to an effect of different satellite fractions above and below each sequence. We use the full simulations instead of mock catalogs to eliminate any possible  contributions from effects due to the  limitations of the  observational galaxy survey data.
Our aim is to compare the normalized 3D clustering amplitudes of galaxy samples with different sSFRs, such that any variation in amplitude is purely a prediction of the model and not influenced by added observational effects.

Figure~\ref{fig04:xi_norm} shows the normalized two-point correlation functions (here in three spatial dimensions, not projected onto the plane of the sky) of galaxy samples selected from the full outputs of the \UM model.
These samples use stellar mass and sSFR cuts identical to those that define the mock galaxy samples used in previous sections, but these samples are \emph{not} subject to the observational effects applied to the simulation to create mock galaxy catalogs that mimic observational data (see \S\ref{subsec04:sims_mocks}).

The left column of Figure~\ref{fig04:xi_norm} shows the distribution of these galaxy samples in the stellar mass--sSFR plane, as in Figures~\ref{fig04:samples_z0p0_obs} and \ref{fig04:samples_z0p45_obs} above. The remaining panels in each row show \xir for each sample divided by \xir of all galaxies \emph{of the relevant type} (centrals, satellites, or both types together), which we call $\xi_{\rm norm}$. We measure \xir instead of \wprp for these galaxy samples because they do not have added line-of-sight position uncertainties.

As seen in Figure~\ref{fig04:xi_norm}, the normalized clustering amplitude decreases smoothly with increasing sSFR, both \emph{within} the star-forming and quiescent populations, and \emph{across} the entire galaxy population. This trend is present both for all galaxies (centrals and satellites) and for centrals only. A deviation from this trend is seen for star-forming satellite galaxies, where the trend is reversed at both $z=0$ and $z=0.45$:\ star-forming satellites with above average sSFR are \emph{more} clustered than those with below average sSFR. This is true for star-forming satellites on both one-halo and two-halo scales, although the difference in the one-halo term is stronger.

To quantify the results of Figure~\ref{fig04:xi_norm} we compute the two-halo intra-sequence relative bias (ISRB) of star-forming and of quiescent mock galaxies, $b_{\rm rel}$, as the mean of $\sqrt{\xi_1(r)/\xi_2(r)}$ over \twohalo, where $\xi_1(r)$ and $\xi_2(r)$ are either for the light blue and dark blue ``main sequence split" samples, respectively, or the red and light red ``main sequence split" samples, respectively, as shown in the figure.
We measure $b_{\rm rel}$ separately for all mock galaxies, mock centrals only, and mock satellites only.

Within the quiescent population the ISRB at $z=0$ is 1.65 for all galaxies, 1.29 for centrals, and 1.17 for satellites. At $z=0.45$ the values are lower but the overall trend is the same:\ quiescent ISRB is greatest for all galaxies (1.42), lower for centrals (1.20), and lowest for satellites (1.10)
We emphasize that the two ``main sequence split" samples within each population (star-forming and quiescent) have the same stellar mass distribution by design; clustering amplitude differences are due exclusively to different sample sSFRs.

For star-forming mock galaxies at $z=0$ the ISRB is 1.29 for all galaxies, 1.19 for centrals, and 0.93 for satellites. The latter value is less than unity, reflecting the greater clustering of star-forming satellites \emph{above} the SFMS compared to those below.
Similarly, at $z=0.45$ the ISRB is 1.10 for all star-forming mock galaxies, 1.09 for centrals, and 0.90 for satellites. While not shown in Figure~\ref{fig04:xi_norm}, the trends at $z=0.9$ again follow those at lower redshift.
These measurements clearly show that the \UM model predicts an anticorrelation between clustering amplitude and sSFR for \emph{central} galaxies, similar to what is seen for all galaxies.
Further, the reversal of the trend for star-forming satellites ($b_{\rm rel} < 1$) means that centrals contribute substantially to the ISRB that is seen for all galaxies.

\section{Summary and Conclusions}\label{sec04:summary}

In this paper we present new measurements of the clustering of stellar mass-complete samples of SDSS galaxies at $z\sim0.03$ as a joint function of stellar mass and sSFR.
We split star-forming galaxies into samples above and below the star-forming main sequence (SFMS) with equivalent stellar mass distributions, and likewise for quiescent galaxies above and below the quiescent sequence.
We compare our SDSS clustering results and C17's comparable measurements at intermediate redshift to mock galaxy catalogs at $z=0$, $0.45$, and $0.9$ based on the empirical \UM galaxy evolution model of \citet{behroozi_etal19}, focusing on the relative bias of various galaxy samples with different sSFR and stellar mass ratios.
We show that this model fits PRIMUS and DEEP2 clustering data well, but the agreement with SDSS is not as strong, particularly within the star-forming population.
Our primary results are:
\begin{enumerate}
\item
Galaxy clustering is a stronger function of sSFR at fixed stellar mass than of stellar mass at fixed sSFR.
Galaxies above the SFMS (with higher sSFR) are less clustered than those below the SFMS (with lower sSFR), at a given stellar mass. We refer to this as \emph{intra-sequence relative bias} (ISRB). A similar trend is present within the quiescent galaxy population.
This result has been shown at $z\sim0.7$ by C17 and we demonstrate here that it is also true at $z\sim0$.
This shows that the scatter observed in the main sequence corresponds to a physical property of the larger-scale environment, in that there are distinct clustering properties for galaxies above and below the sequence. A similar correlation is likewise present within the quiescent population. 
\item
Tests with mock catalogs from the \UM suggest that central galaxies are the driver of our result that sSFR ratio (and not stellar mass ratio) is the key factor influencing the clustering differences between galaxy samples. While this effect is also present for satellite galaxies, the correlation between sSFR and clustering amplitude is stronger for centrals in the mock catalogs. This is consistent with central galaxy assembly bias, and/or distinct stellar-to-halo mass relations for star-forming and quiescent (central) galaxies.
\item
The empirical model of \citet{behroozi_etal19} fits combined PRIMUS and DEEP2 clustering data well at intermediate redshift ($z\sim0.45$ and $z\sim0.9$), i.e., galaxy bias increases smoothly with decreasing sSFR in the model as well as in the data.
At low redshift ($z\sim0$), the model does not reproduce SDSS data well in terms of ISRB within the star-forming population. We show that increasing the correlation between galaxy SFR and halo accretion rate in the model improves the agreement with the observations.
\item
Measurements of galaxy bias as a function of sSFR, and of relative bias versus sSFR ratio for different galaxy samples, are highly constraining for models of galaxy evolution. Forward modeling with mock galaxy catalogs based on empirical models, as performed here, allows for comparisons of data and models at intermediate redshift without the need for stellar mass-complete galaxy samples, which are currently restricted to relatively high mass at such redshifts.
\end{enumerate}

Findings (1) and (2) above are in agreement with \citet{rodriguez-puebla_etal15}, who find a statistically significant difference between the SHMRs of red and blue central galaxies in the SDSS at halo masses of $\sim10^{12}$ \msun. These results are also consistent with \citet{matthee_schaye19}, who measure the joint stellar mass and redshift dependence of the scatter in the SFMS using the EAGLE simulation, concluding that the scatter in the SFMS at $z=0.1$ results from variations in halo formation times, consistent with galaxy assembly bias.
In contrast, \citet[][]{odonnell_etal20} estimate halo accretion rates and isolated galaxy sSFRs in both the SDSS at $z<0.123$ and \UM, and find no statistically significant correlation between halo assembly history and sSFR. However, their study is limited to $10.5<\logm<11.0$ and does not subdivide galaxies \emph{within} the main sequence as we do here. \citet[][]{odonnell_etal20} also use satellite galaxies to probe dark matter accretion; however, satellites may be a biased tracer of the dark matter distribution, which would in turn impact inferences about dark matter accretion.

Our new measurements anticipate the arrival of near-future galaxy redshift surveys that will provide unprecedented statistical studies of large-scale structure at higher redshift. A growing body of literature indicates that considerable additional information about structure growth and the galaxy-halo connection may be contained in environmental statistics beyond the standard two-point function measurements \citep[e.g.][]{wang_etal19, banerjee_abel20, uhlemann_etal20}. The results presented here offer an explicit  demonstration that the same is true of the two-point function itself: by dividing galaxy samples into two-dimensional subsamples based on stellar mass and sSFR, particularly when including subsamples {\it within} each of the star-forming and quiescent populations, clustering measurements of the subsamples can be mined for valuable additional constraints on galaxy formation beyond what can be achieved with standard clustering measurements split on sSFR.

\acknowledgements
AMB and ALC acknowledge support from the Ingrid and Joseph W.\ Hibben endowed chair at UC San Diego.
Work done at Argonne National Laboratory was supported by the U.S. Department of Energy, Office of Science,  Office  of  Nuclear  Physics,  under  contract  DE-AC02-06CH11357.  We  gratefully  acknowledge  use  of  the  Bebop cluster  in  the  Laboratory  Computing  Resource  Center  at Argonne National Laboratory. Computational work for this paper was also performed on the Phoenix cluster at Argonne National Laboratory, jointly maintained by the Cosmological Physics and Advanced Computing (CPAC) group and by the Computing, Environment, and Life Sciences (CELS) directorate.  PB was partially funded by a Packard Fellowship, Grant \#2019-69646.

Funding for SDSS-III has been provided by the Alfred P.\ Sloan Foundation, the Participating Institutions, the National Science Foundation, and the U.S.\ Department of Energy Office of Science. The SDSS-III website is http://www.sdss3.org/.
SDSS-III is managed by the Astrophysical Research Consortium for the Participating Institutions of the SDSS-III Collaboration including the University of Arizona, the Brazilian Participation Group, Brookhaven National Laboratory, Carnegie Mellon University, University of Florida, the French Participation Group, the German Participation Group, Harvard University, the Instituto de Astrofisica de Canarias, the Michigan State/Notre Dame/JINA Participation Group, Johns Hopkins University, Lawrence Berkeley National Laboratory, Max Planck Institute for Astrophysics, Max Planck Institute for Extraterrestrial Physics, New Mexico State University, New York University, Ohio State University, Pennsylvania State University, University of Portsmouth, Princeton University, the Spanish Participation Group, University of Tokyo, University of Utah, Vanderbilt University, University of Virginia, University of Washington, and Yale University.

The CosmoSim database used in this paper is a service by the Leibniz-Institute for Astrophysics Potsdam (AIP). The MultiDark database was developed in cooperation with the Spanish MultiDark Consolider Project CSD2009-00064.
The authors gratefully acknowledge the Gauss Centre for Supercomputing e.V.\ (www.gauss-centre.eu) and the Partnership for Advanced Supercomputing in Europe (PRACE, www.prace-ri.eu) for funding the MultiDark simulation project by providing computing time on the GCS Supercomputer SuperMUC at Leibniz Supercomputing Centre (LRZ, www.lrz.de).
The Bolshoi simulations have been performed within the Bolshoi project of the University of California High-Performance AstroComputing Center (UC-HiPACC) and were run at the NASA Ames Research Center.

\bibliography{refs}

\section*{Appendix:\ Uniform Stellar Mass Complete SDSS Galaxy Samples}

Here we present measurements of \wprp for SDSS galaxy samples that use a single stellar mass completeness cut. These samples are nearly identical to the samples described in \S\ref{subsubsec04:samples_sdss} and used for the results presented in \S\ref{sec04:wp_bias}, but are limited to $\logm > 9.75$ for both star-forming and quiescent galaxies so as to be better suited for comparisons with theoretical models.

As in the main text, the four sets of galaxy samples are (1) ``star-forming/quiescent split," in which galaxies are divided into star-forming and quiescent samples by Eq.~\ref{eq04:sdss_ssfr_cut}; (2) ``main sequence split," in which the star-forming and quiescent populations are split into samples above and below each population's main sequence; (3) ``sSFR cuts," in which galaxies are split into six bins in sSFR; and (4) ``stellar mass/sSFR grid," in which galaxies are divided into seven samples using two bins in stellar mass and four bins in sSFR.

\begin{figure*}[ht]
\centering
\includegraphics[width=\linewidth]{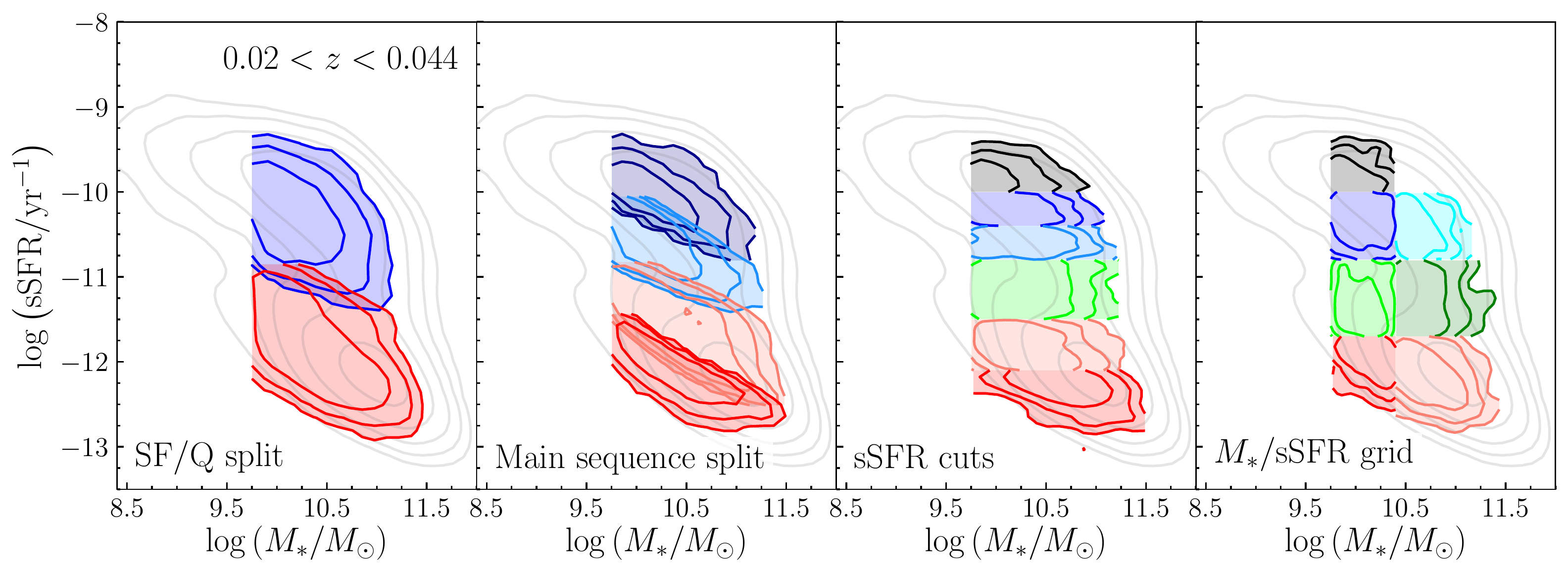}
\includegraphics[width=\linewidth]{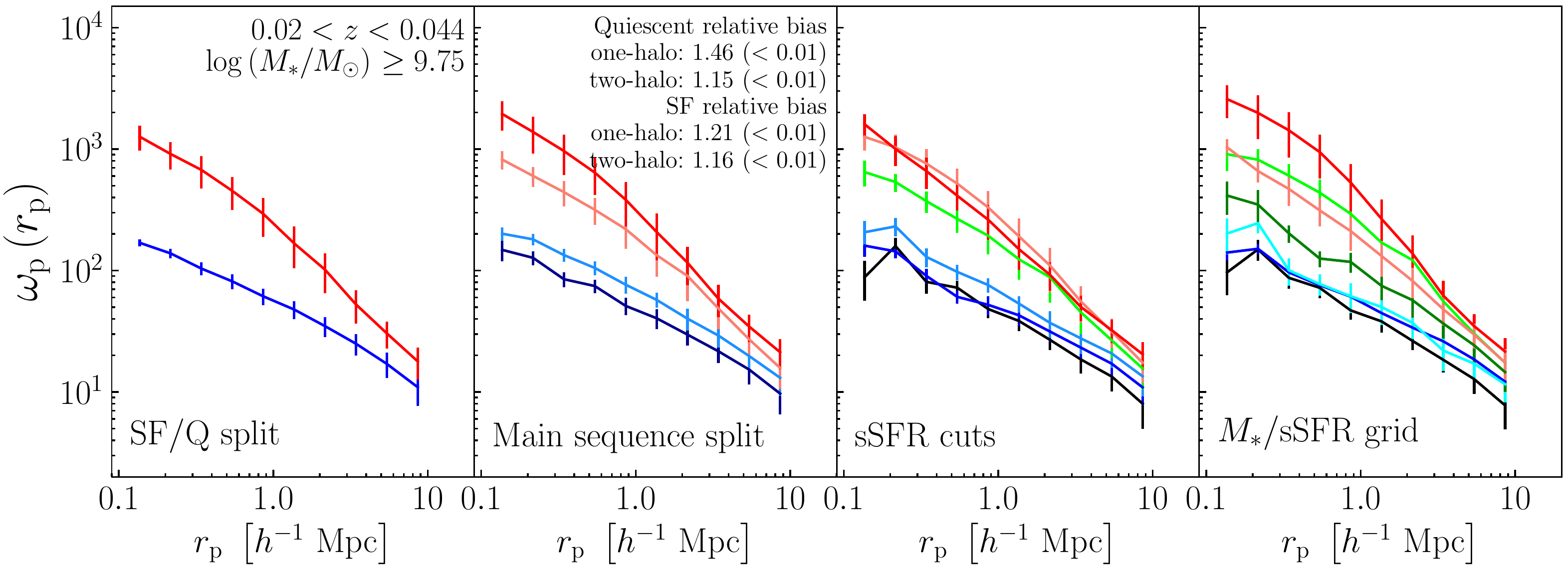}
\caption{
Same as Figure~\ref{fig04:sdss_samples_sm9p75q_9p25sf} but using a stellar mass limit of ${\logm \ge 9.75}$ for all galaxy samples.
}
\label{fig04:sdss_samples_sm9p75}
\end{figure*}

The top row of Figure~\ref{fig04:sdss_samples_sm9p75} shows the distribution of all galaxy samples in the stellar mass--sSFR plane; \wprp for each sample is shown in the bottom row.
Table~\ref{tab04:samples_sdss_strict} lists the stellar mass and sSFR cuts used for each sample, and Table~\ref{tab04:wprp_sdss_strict} contains \wprp for each sample for 10 logarithmic bins in \rp between $\log(\rp/(\Mpch))=-0.8$ and $\log(\rp/(\Mpch))=1.2$.

\begin{deluxetable*}{llrrrrrrr}[bp]
\tablecaption{Strict stellar mass complete SDSS galaxy samples ${(\logm \geq 9.75)}$. All samples span $0.02 < z < 0.0435$ and have median redshift $z_{\rm med}\simeq0.033$.
\label{tab04:samples_sdss_strict}
}

\tablehead{
\colhead{\multirow{2}{*}{Run}} &
\colhead{\multirow{2}{*}{Sample}} &
\colhead{\multirow{2}{*}{$N_{\rm gal}$}} &
\multicolumn{3}{c}{\logm} &
\multicolumn{3}{c}{$\log($sSFR/yr$^{-1})$} \\
\colhead{} & \colhead{} & \colhead{} &
\colhead{min} & \colhead{mean} & \colhead{max} &
\colhead{min} & \colhead{mean} & \colhead{max}
}

\startdata
${\rm SF/Q\ split}$	& \textcolor{blue}{\textbf{blue}}	& 13548	& 9.75	& 10.20	& 11.98	& -11.64	& -10.30	& -8.41 \\ 
	& \textcolor{red}{\textbf{red}}	& 13449	& 9.75	& 10.42	& 12.42	& -13.40	& -11.87	& -10.80 \\ 
${\rm Main\ sequence\ split}$	& \textcolor{darkblue}{\textbf{dark blue}}	& 6762	& 9.75	& 10.20	& 11.35	& -11.20	& -10.03	& -8.41 \\ 
	& \textcolor{dodgerblue}{\textbf{light blue}}	& 6779	& 9.75	& 10.20	& 11.43	& -11.54	& -10.57	& -10.05 \\ 
	& \textcolor{salmon}{\textbf{light red}}	& 6712	& 9.75	& 10.42	& 11.59	& -12.70	& -11.60	& -10.80 \\ 
	& \textcolor{red}{\textbf{red}}	& 6732	& 9.75	& 10.42	& 11.75	& -13.40	& -12.14	& -11.40 \\ 
${\rm sSFR\ cuts}$	& \textbf{black}	& 3109	& 9.75	& 10.03	& 11.15	& -10.00	& -9.82	& -9.01 \\ 
	& \textcolor{blue}{\textbf{blue}}	& 5272	& 9.75	& 10.15	& 11.24	& -10.40	& -10.20	& -10.00 \\ 
	& \textcolor{dodgerblue}{\textbf{light blue}}	& 3658	& 9.75	& 10.28	& 11.35	& -10.80	& -10.58	& -10.40 \\ 
	& \textcolor{lime}{\textbf{light green}}	& 4342	& 9.75	& 10.27	& 11.57	& -11.50	& -11.16	& -10.80 \\ 
	& \textcolor{salmon}{\textbf{light red}}	& 6134	& 9.75	& 10.36	& 11.55	& -12.10	& -11.83	& -11.50 \\ 
	& \textcolor{red}{\textbf{red}}	& 4462	& 9.76	& 10.70	& 12.00	& -13.40	& -12.32	& -12.10 \\ 
$M_*/{\rm sSFR\ grid}$	& \textbf{black}	& 2846	& 9.75	& 9.98	& 10.40	& -10.00	& -9.82	& -8.41 \\ 
	& \textcolor{blue}{\textbf{blue}}	& 6668	& 9.75	& 10.06	& 10.40	& -10.80	& -10.33	& -10.00 \\ 
	& \textcolor{lime}{\textbf{light green}}	& 3940	& 9.75	& 10.06	& 10.40	& -11.70	& -11.29	& -10.80 \\ 
	& \textcolor{red}{\textbf{red}}	& 3104	& 9.75	& 10.15	& 10.40	& -13.40	& -11.97	& -11.70 \\ 
	& \textcolor{cyan}{\textbf{cyan}}	& 2262	& 10.40	& 10.61	& 11.35	& -10.80	& -10.44	& -10.00 \\ 
	& \textcolor{green}{\textbf{dark green}}	& 2037	& 10.40	& 10.68	& 11.43	& -11.70	& -11.26	& -10.80 \\ 
	& \textcolor{salmon}{\textbf{light red}}	& 5842	& 10.40	& 10.75	& 11.50	& -13.36	& -12.19	& -11.70 \\ 
\enddata
\end{deluxetable*}

\begin{deluxetable*}{rrrrrrrr}[tb]
\tablecaption{\wprp\tablenotemark{a} for strict stellar mass complete SDSS galaxy samples ${(\logm \geq 9.75)}$.
\label{tab04:wprp_sdss_strict}
}

\tablehead{
\multicolumn{3}{c}{${\rm Star}$-${\rm forming/quiescent\ split}$} &
\multicolumn{5}{c}{${\rm Main\ sequence\ split}$} \\
\multicolumn{8}{c}{\vspace{-8pt}} \\
\cline{1-8}
\multicolumn{8}{c}{\vspace{-5pt}} \\
\multicolumn{1}{r}{\rp\tablenotemark{b}\phantom{0}} &
\colhead{\textcolor{blue}{\textbf{blue}}}	&
\colhead{\textcolor{red}{\textbf{red}}}	&
\multicolumn{1}{r}{\rp\phantom{0}} &
\colhead{\textcolor{darkblue}{\textbf{dark blue}}}	&
\colhead{\textcolor{dodgerblue}{\textbf{light blue}}}	&
\colhead{\textcolor{salmon}{\textbf{light red}}}	&
\colhead{\textcolor{red}{\textbf{red}}}
}

\startdata
$0.136$	& $169.2\ (4.5)$	& $1264.9\ (38.9)$	& $0.136$	& $147.9\ (4.4)$	& $201.1\ (8.2)$	& $820.0\ (22.3)$	& $1947.3\ (69.9)$ \\
$0.216$	& $138.4\ (3.5)$	& $911.8\ (32.0)$	& $0.216$	& $127.3\ (3.7)$	& $180.6\ (5.2)$	& $601.2\ (19.5)$	& $1383.6\ (66.3)$ \\
$0.343$	& $103.8\ (3.4)$	& $673.3\ (27.4)$	& $0.343$	& $84.9\ (3.1)$	& $134.1\ (4.5)$	& $442.1\ (17.0)$	& $964.3\ (54.9)$ \\
$0.543$	& $81.4\ (2.7)$	& $452.7\ (19.7)$	& $0.543$	& $74.5\ (2.4)$	& $105.0\ (3.6)$	& $316.2\ (14.3)$	& $640.2\ (33.0)$ \\
$0.861$	& $61.3\ (2.4)$	& $293.0\ (16.0)$	& $0.861$	& $51.2\ (2.2)$	& $76.4\ (2.9)$	& $219.2\ (12.2)$	& $382.7\ (25.1)$ \\
$1.364$	& $47.8\ (2.0)$	& $167.4\ (9.9)$	& $1.364$	& $40.5\ (1.9)$	& $57.4\ (2.0)$	& $133.9\ (7.8)$	& $207.8\ (14.8)$ \\
$2.162$	& $34.9\ (1.6)$	& $101.7\ (5.8)$	& $2.162$	& $29.7\ (1.5)$	& $40.2\ (1.8)$	& $90.2\ (5.9)$	& $116.0\ (6.7)$ \\
$3.426$	& $25.0\ (1.3)$	& $52.6\ (2.3)$	& $3.426$	& $21.6\ (1.2)$	& $29.2\ (1.4)$	& $48.7\ (2.5)$	& $58.7\ (2.7)$ \\
$5.431$	& $17.0\ (1.1)$	& $30.4\ (1.7)$	& $5.431$	& $15.3\ (1.1)$	& $19.7\ (1.1)$	& $27.1\ (1.5)$	& $34.5\ (1.9)$ \\
$8.607$	& $11.0\ (0.9)$	& $17.9\ (1.3)$	& $8.607$	& $9.7\ (1.0)$	& $13.1\ (0.9)$	& $15.8\ (1.3)$	& $21.2\ (1.3)$ \\
\cutinhead{${\rm sSFR\ cuts}$}
\multicolumn{1}{r}{\rp\phantom{0}} &
\colhead{\textbf{black}}	&
\colhead{\textcolor{blue}{\textbf{blue}}}	&
\colhead{\textcolor{dodgerblue}{\textbf{light blue}}}	&
\colhead{\textcolor{lime}{\textbf{light green}}}	&
\colhead{\textcolor{salmon}{\textbf{light red}}}	&
\colhead{\textcolor{red}{\textbf{red}}} & \\
\multicolumn{8}{c}{\vspace{-8pt}} \\
\cline{1-8}
\multicolumn{8}{c}{\vspace{-5pt}} \\
$0.136$	& $88.0\ (7.3)$	& $160.3\ (6.1)$	& $206.7\ (14.3)$	& $646.2\ (30.6)$	& $1261.0\ (34.2)$	& $1599.6\ (50.1)$	&  \\
$0.216$	& $159.1\ (5.7)$	& $143.8\ (4.2)$	& $230.9\ (10.7)$	& $535.4\ (19.5)$	& $1033.3\ (39.1)$	& $1002.8\ (48.8)$	&  \\
$0.343$	& $80.7\ (3.9)$	& $90.3\ (3.6)$	& $129.5\ (5.7)$	& $372.2\ (18.4)$	& $760.2\ (33.9)$	& $658.8\ (32.0)$	&  \\
$0.543$	& $72.2\ (3.0)$	& $60.7\ (2.0)$	& $97.4\ (3.8)$	& $266.2\ (13.7)$	& $519.7\ (26.2)$	& $412.7\ (19.3)$	&  \\
$0.861$	& $48.2\ (2.2)$	& $52.2\ (2.0)$	& $76.0\ (2.8)$	& $193.4\ (11.0)$	& $330.6\ (19.5)$	& $262.0\ (14.3)$	&  \\
$1.364$	& $38.4\ (2.1)$	& $42.7\ (1.8)$	& $53.3\ (1.9)$	& $123.4\ (8.2)$	& $190.7\ (13.5)$	& $149.6\ (7.6)$	&  \\
$2.162$	& $26.9\ (1.4)$	& $31.5\ (1.5)$	& $37.1\ (1.7)$	& $88.1\ (6.5)$	& $110.1\ (7.0)$	& $92.2\ (4.2)$	&  \\
$3.426$	& $18.5\ (1.1)$	& $23.2\ (1.2)$	& $27.6\ (1.4)$	& $45.3\ (2.7)$	& $55.7\ (2.6)$	& $50.1\ (2.1)$	&  \\
$5.431$	& $13.4\ (1.0)$	& $17.1\ (1.1)$	& $20.7\ (1.2)$	& $26.6\ (1.7)$	& $31.5\ (1.8)$	& $32.3\ (1.7)$	&  \\
$8.607$	& $8.0\ (1.0)$	& $10.9\ (0.9)$	& $13.5\ (1.0)$	& $15.6\ (1.3)$	& $17.4\ (1.3)$	& $20.5\ (1.3)$	&  \\
\cutinhead{${\rm Stellar\ mass/sSFR\ grid}$}
\multicolumn{1}{r}{\rp\phantom{0}} &
\colhead{\textbf{black}}	&
\colhead{\textcolor{blue}{\textbf{blue}}}	&
\colhead{\textcolor{lime}{\textbf{light green}}}	&
\colhead{\textcolor{red}{\textbf{red}}}	&
\colhead{\textcolor{cyan}{\textbf{cyan}}}	&
\colhead{\textcolor{green}{\textbf{dark green}}}	&
\colhead{\textcolor{salmon}{\textbf{light red}}} \\
\multicolumn{8}{c}{\vspace{-8pt}} \\
\cline{1-8}
\multicolumn{8}{c}{\vspace{-5pt}} \\
$0.136$	& $96.5\ (8.9)$	& $140.6\ (5.6)$	& $906.9\ (35.3)$	& $2577.9\ (138.0)$	& $201.7\ (16.1)$	& $414.5\ (29.5)$	& $1035.2\ (32.7)$ \\
$0.216$	& $149.3\ (5.6)$	& $151.0\ (4.6)$	& $817.3\ (31.7)$	& $1986.5\ (128.9)$	& $245.6\ (11.6)$	& $349.0\ (32.2)$	& $659.0\ (21.5)$ \\
$0.343$	& $87.1\ (4.3)$	& $96.6\ (3.2)$	& $603.9\ (28.5)$	& $1427.8\ (91.6)$	& $100.4\ (5.4)$	& $202.5\ (8.5)$	& $469.5\ (19.4)$ \\
$0.543$	& $72.1\ (3.0)$	& $75.7\ (2.5)$	& $437.6\ (22.3)$	& $944.5\ (55.1)$	& $77.4\ (3.7)$	& $125.5\ (4.5)$	& $312.6\ (12.4)$ \\
$0.861$	& $47.0\ (2.3)$	& $61.0\ (2.2)$	& $293.2\ (16.4)$	& $531.0\ (35.7)$	& $61.4\ (3.9)$	& $117.8\ (4.6)$	& $211.3\ (10.8)$ \\
$1.364$	& $38.1\ (2.1)$	& $44.7\ (1.7)$	& $169.7\ (10.8)$	& $266.3\ (21.1)$	& $50.1\ (3.3)$	& $74.8\ (4.3)$	& $131.7\ (6.8)$ \\
$2.162$	& $26.4\ (1.3)$	& $33.9\ (1.4)$	& $121.9\ (9.1)$	& $138.4\ (10.4)$	& $37.1\ (2.3)$	& $57.1\ (3.3)$	& $82.3\ (3.8)$ \\
$3.426$	& $18.4\ (1.1)$	& $26.1\ (1.2)$	& $55.8\ (3.0)$	& $62.1\ (3.1)$	& $21.9\ (1.8)$	& $36.6\ (2.1)$	& $48.0\ (2.1)$ \\
$5.431$	& $12.8\ (0.9)$	& $18.5\ (1.0)$	& $30.2\ (1.6)$	& $34.8\ (1.9)$	& $17.1\ (1.2)$	& $24.2\ (1.8)$	& $29.6\ (1.6)$ \\
$8.607$	& $7.8\ (0.9)$	& $12.1\ (0.9)$	& $17.4\ (1.3)$	& $21.5\ (1.4)$	& $11.6\ (1.0)$	& $14.6\ (1.3)$	& $17.5\ (1.2)$ \\
\enddata
\tablenotetext{a}{Units of \wprp are \Mpch with $h=0.7$. Sample parameters are presented in Table~\ref{tab04:samples_sdss_strict}.}
\tablenotetext{b}{Shown are bin centers in units of \Mpch with $h=0.7$ for 10 logarithmic bins between $-0.8$ and $1.2$.}
\end{deluxetable*}

\end{document}